\documentclass[useAMS,usenatbib]{mnras}
\usepackage{amsmath}
\usepackage{amssymb}
\usepackage{graphicx}
\usepackage{times}
\usepackage{hyperref}
\usepackage[normalem]{ulem}

\newcommand\textlcsc[1]{\textsc{\MakeLowercase{#1}}}

\newcommand{\Hmocks}{\textsc{hits-mocks}}
\newcommand{\Dmocks}{\textsc{icc-mocks}}
\newcommand{\Dmocksnoex}{\textsc{icc-mocks-noex}}

\newcommand{\gaia}{\emph{Gaia}}

\newcommand{\enbid}{\textsc{enbid}}

\def\jref@jnl#1{{\rm#1}}

\def\aj{\jref@jnl{AJ}}                   
\def\araa{\jref@jnl{ARA\&A}}             
\def\apj{\jref@jnl{ApJ}}                 
\def\apjl{\jref@jnl{ApJ}}                
\def\apjs{\jref@jnl{ApJS}}               
\def\ao{\jref@jnl{Appl.~Opt.}}           
\def\apss{\jref@jnl{Ap\&SS}}             
\def\aap{\jref@jnl{A\&A}}                
\def\aapr{\jref@jnl{A\&A~Rev.}}          
\def\aaps{\jref@jnl{A\&AS}}              
\def\azh{\jref@jnl{AZh}}                 
\def\baas{\jref@jnl{BAAS}}               
\def\jrasc{\jref@jnl{JRASC}}             
\def\memras{\jref@jnl{MmRAS}}            
\def\mnras{\jref@jnl{MNRAS}}             
\def\pasa{\jref@jnl{PASA}}
\def\pra{\jref@jnl{Phys.~Rev.~A}}        
\def\prb{\jref@jnl{Phys.~Rev.~B}}        
\def\prc{\jref@jnl{Phys.~Rev.~C}}        
\def\prd{\jref@jnl{Phys.~Rev.~D}}        
\def\pre{\jref@jnl{Phys.~Rev.~E}}        
\def\prl{\jref@jnl{Phys.~Rev.~Lett.}}    
\def\pasp{\jref@jnl{PASP}}               
\def\pasj{\jref@jnl{PASJ}}               
\def\qjras{\jref@jnl{QJRAS}}             
\def\skytel{\jref@jnl{S\&T}}             
\def\solphys{\jref@jnl{Sol.~Phys.}}      
\def\sovast{\jref@jnl{Soviet~Ast.}}      
\def\ssr{\jref@jnl{Space~Sci.~Rev.}}     
\def\zap{\jref@jnl{ZAp}}                 
\def\nat{\jref@jnl{Nature}}              
\def\iaucirc{\jref@jnl{IAU~Circ.}}       
\def\aplett{\jref@jnl{Astrophys.~Lett.}} 
\def\apspr{\jref@jnl{Astrophys.~Space~Phys.~Res.}}
\def\bain{\jref@jnl{Bull.~Astron.~Inst.~Netherlands}} 
\def\fcp{\jref@jnl{Fund.~Cosmic~Phys.}}  
\def\gca{\jref@jnl{Geochim.~Cosmochim.~Acta}}   
\def\grl{\jref@jnl{Geophys.~Res.~Lett.}} 
\def\jcp{\jref@jnl{J.~Chem.~Phys.}}      
\def\jgr{\jref@jnl{J.~Geophys.~Res.}}    
\def\jqsrt{\jref@jnl{J.~Quant.~Spec.~Radiat.~Transf.}}
\def\memsai{\jref@jnl{Mem.~Soc.~Astron.~Italiana}}
\def\nphysa{\jref@jnl{Nucl.~Phys.~A}}   
\def\physrep{\jref@jnl{Phys.~Rep.}}   
\def\physscr{\jref@jnl{Phys.~Scr}}   
\def\planss{\jref@jnl{Planet.~Space~Sci.}}   
\def\procspie{\jref@jnl{Proc.~SPIE}}   

\usepackage{datatool}   

\title[\textit{Aurigaia}: Cosmological Gaia mocks]{\textit{Aurigaia}: mock \emph{Gaia} DR2 stellar catalogues from the Auriga cosmological simulations}
\author[Grand et al.]{\parbox[t]{\textwidth}{
Robert J. J. Grand$^{1,2}$\thanks{robert.grand@h-its.org}, John Helly$^3$, Azadeh Fattahi$^{3}$, Marius Cautun$^{3}$, Shaun Cole$^3$, Andrew P. Cooper$^3$, Alis J. Deason$^{3}$, Carlos Frenk$^3$, Facundo A. G\'{o}mez$^{4,5,6}$, Jason A. S. Hunt$^{7}$, Federico Marinacci$^{8,9}$, R\"{u}diger Pakmor$^1$, Christine M. Simpson$^1$, Volker Springel$^{1,2,6}$, Dandan Xu$^1$} \vspace{10pt} \\
$^1$Heidelberger Institut f\"{u}r Theoretische Studien, Schloss-Wolfsbrunnenweg 35, 69118 Heidelberg, Germany\\
$^2$Zentrum f\"{u}r Astronomie der Universit\"{a}t Heidelberg, Astronomisches Recheninstitut, M\"{o}nchhofstr. 12-14, 69120 Heidelberg, Germany\\
$^3$Institute for Computational Cosmology, Department of Physics, Durham University, South Road Durham DH1 3LE, UK \\
$^{4}$Instituto de Investigaci{\'o}n Multidisciplinar en Ciencia yTecnolog{\'i}a, Universidad de La Serena, Ra{\'u}l Bitr{\'a}n 1305, La Serena, Chile\\
$^{5}$Departamento de F{\'i}sica y Astronom{\'i}a, Universidad de LaSerena, Av. Juan Cisternas 1200 N, La Serena, Chile\\
$^6$Max-Planck-Institut f\"{u}r Astrophysik, Karl-Schwarzschild-Str. 1, D-85748, Garching, Germany  \\
$^7$Dunlap Institute for Astronomy and Astrophysics, University of Toronto, 50 St. George Street, Toronto, Ontario, M5S 3H4, Canada\\
$^8$Department of Physics, Kavli Institute for Astrophysics and Space Research, MIT, Cambridge, MA 02139, USA \\
$^9$Harvard-Smithsonian Center for Astrophysics, 60 Garden Street, Cambridge, MA 02138, USA
}

\pubyear{2018}
    
\hypersetup{draft} 
\begin{document}

\label{firstpage}

\pagerange{\pageref{firstpage}--\pageref{lastpage}}
\maketitle

\begin{abstract}
  We present and analyse mock stellar catalogues that match the selection criteria and
  observables (including uncertainties) of the \emph{Gaia} satellite
  data release 2 (DR2). The source are six cosmological high-resolution magneto-hydrodynamic
  $\Lambda$CDM zoom simulations of the formation of Milky Way analogues from the \textlcsc{AURIGA} project. Mock data are provided for stars with $V < 16$ mag,
  and $V < 20$ mag at $|b|>20$ degrees. The mock catalogues are made
  using two different methods: the public \textlcsc{SNAPDRAGONS} code,
  and a method based on that of Lowing et al. that preserves the
  phase-space distribution of the model stars. These publicly available catalogues contain
  5-parameter astrometry, radial velocities, multi-band photometry,
  stellar parameters, dust extinction values, and uncertainties in all
  these quantities. In addition, we provide the gravitational potential and
  information on the origin of each star. By way of demonstration, we apply the mock catalogues to analyses of the young stellar
  disc and the stellar halo. We show that: i)~the young
  outer stellar disc exhibits a flared distribution that is detectable
  in the height and vertical velocity distribution of $\rm A$- and
  $\rm B$-dwarf stars up to radii of $\sim 15$ kpc; and
  ii)~the spin of the stellar halo out to 100~kpc can be accurately measured with
  \emph{Gaia} DR2 RR Lyrae stars. These catalogues are well suited for comparisons with observations
  and should help to: i)~develop and test analysis methods for the
  \emph{Gaia} DR2 data; ii)~gauge the limitations and biases of the
    data and iii)~ interpret the data in the light of theoretical
    predictions from realistic {\em ab initio} simulations of galaxy
    formation in the $\Lambda$CDM cosmological model. 
\end{abstract}


\section{Introduction}
Over the next five years, our view of the Milky Way galaxy will be revolutionised by the European Space Agency's cornerstone \emph{Gaia} mission \citep{GC+16}, which aims to provide positions and velocities for billions of stars in the Galaxy -- a 10000-fold increase in sample size and 100-fold increase in precision over its predecessor, \emph{Hipparcos} \citep{VLF07}. The second \emph{Gaia} date release \citep[DR2,][]{Gaia18a,Gaia18b,Gaia18c} will already provide astrometric and photometric data in three bands for $\sim 1.4$ billion sources over the entire sky. A fraction of this dataset will contain also measurements for radial velocities, extinction and effective temperatures. With subsequent \emph{Gaia} data releases, in combination with several major current and future spectroscopic surveys, such as SDSS/APOGEE \citep{MSF17}, DESI \citep{DESI16}, Gaia-ESO \citep{GRA12}, LAMOST \citep{CHY12}, GALAH \citep{MSB17} and 4MOST \citep{DJB14}, and asteroseismic surveys, such as K2 \citep{SZE17}, TESS \citep{CSK16} and PLATO \citep{RCA14}, additional data for tens of millions of stars will become available that include chemical abundances, radial velocities, and stellar ages. 

In principle, this huge amount of high-dimensional empirical information about the stellar component of our Galaxy holds the key to unveiling its current state through precise identification of disc, bulge and halo substructure, and its formation history \citep[see][for a recent overview]{RB13R}. Given that the Milky Way is thought to be fairly typical for its mass \citep[although see][]{Bell2017,Cautun2018} within the standard model of cosmology -- the Lambda Cold Dark Matter ($\Lambda$CDM) paradigm -- this multi-dimensional star-by-star information provides a unique window into the formation of $L_*$ galaxies in general, as well as a test of the predictions of $\Lambda$CDM.

This new wealth of observational data is only a partial snapshot of the current distribution of stars in our quadrant of the Milky Way, however, and its interpretation requires some form of modelling. Widely employed modelling techniques include dynamical models such as (quasi-) distribution functions \citep{Bi10,BR13,TBR16}; Torus mapping \citep{BM16}; Made-to-Measure (M2M) models \citep{ST96,Hun13} that aim to characterise the current structure of the major Galactic components; and self-consistent $N$-body models that provide testable predictions for the effects of various evolutionary processes \citep[e.g.][]{GKC12,KGG17,FDH17}. A crucial aspect in the quest to draw reliable conclusions from any of these techniques is to understand the limitations, biases and quality of the observational data. Specifically, the effects of survey selection functions, sample size, survey volume, accuracy of phase space and spectroscopic measurements, dust obscuration and image crowding influence inferences as to the true phase-space distribution of stars.

A pragmatic solution to these problems is to generate and analyse
synthetic Milky Way catalogues cast in the observational frame of the
survey \citep{BS80,RC86,BRC87}.  ``Mock catalogues'' of this general
type were first used in cosmology in the 2000s
\citep[e.g.][]{Cole2005} and have now become an essential tool for the
design and analysis of large galaxy and quasar surveys. Realistic mock
catalogues provide assessments of an instrument's capabilities and
biases, tests of statistical modelling techniques applied to realistic
representations of observational data, and detailed comparisons
between theoretical predictions and observations. Perhaps one of the
best known recent attempts is the Besan\c{c}on model \citep{RRD03},
which provides a disc (or set of discs) with a set of coeval and
isothermal (single velocity dispersion) stellar populations assumed to
be in equilibrium, with analytically specified distributions of
density, metallicity and age. This has been the basis of the
  \emph{Gaia Universe Model} \citep[GUMS;][]{RL12}. However, these
models are not dynamically consistent and oversimplify the structure
of the Galaxy, particularly the stellar halo which is modelled as a
smooth component. An important advance was made by \citet{SBJ11}, who
developed the \textlcsc{Galaxia} code for creating mock stellar
catalogues either analytically or from phase space sampling of hybrid
semi-analytic-$N$-body simulations to represent stellar haloes in a
cosmological context \citep{BJ05,CCF10}. \citet{RDF18} have
  developed a mock catalogue designed specifically for {\it Gaia} DR2
  based on \textlcsc{Galaxia}. Building on the method of
  \citet{SBJ11}, \citet{LWC15} developed a technique to distribute
  synthetic stars sampled from a cosmological $N$-body simulation in
  such a way as to preserve the phase-space properties of their parent
  stellar populations. In a separate method, \citet{HKG15} introduced
  the \textlcsc{SNAPDRAGONS} code that generates a mock catalogue
  taking into account \emph{Gaia} errors and extinction, and
  demonstrated the resulting observable kinematics of stars around a
  spiral arm in an idealized smoothed particle hydrodynamic simulation
  set up in isolation.

One of the goals of modern Galactic astronomy is to compare predictions of \emph{ab initio} cosmological formation models with the high-dimensional observational data provided by Galactic surveys in order to elucidate the evolutionary history of the Galaxy. Mock stellar catalogues based on full hydrodynamical cosmological simulations are an appealing prospect to fulfil this aim. This would provide us with a window into how different types of stars that originate from cosmological initial conditions are distributed in phase space. Given that the details of these distributions will depend on the formation history of the Milky Way, multiple mock catalogues derived from simulations that span a range of formation histories will be desirable for many aspects of disc and halo formation. 

Until recently, the availability of realistic cosmological simulations of Milky Way analogues has been limited due to a combination of numerical hindrances and insufficiently realistic astrophysical modelling of important physical effects, such as feedback processes \citep{KG91,NS00,GWB10,SWS11}. This situation has improved and cosmological zoom simulations have now become sophisticated enough to produce sets of high-resolution Milky Way analogues in statistically meaningful numbers \citep[e.g.][]{MPS14,WDS15,FNS16,GKH18}. In particular, the \textlcsc{AURIGA} simulation suite \citep{GGM17} consists of 40 Milky Way mass haloes simulated at resolutions comparable to the most modern idealised simulations ($6\times10^3$ to $5\times10^4$ $\rm M_{\odot}$ per baryonic element) with a comprehensive galaxy formation model, including physical processes such as magnetic fields \citep{PMS14} and feedback from active galactic nuclei \citep{SMH05} and stars \citep{VGS13}. These simulations have been shown to produce disc-dominated, star-forming late-type spiral galaxies that are broadly consistent with a plethora of observational data such as star formation histories, abundance matching predictions, gas fractions, sizes, and rotation curves of $L_{*}$ galaxies \citep{GGM17}. Furthermore, they are sufficiently detailed to address questions related to chemodynamic properties of the Milky Way, such as the origin of the chemical thin-thick disc dichotomy \citep{GBG18}, the formation of bars, spiral arms and warps \citep{GWG16}, and the properties of the stellar halo \citep[][]{MGG16,MGG18} and satellite galaxies \citep{SGG17}. The confluence of these advanced simulation techniques with the new \emph{Gaia} and ground-based data will transform, at a fundamental level, the understanding of our Galaxy in its cosmological context.

The aim of this paper is to present two sets of mock \emph{Gaia} DR2 stellar catalogues generated from the \textlcsc{AURIGA} cosmological simulations: one set generated with a parallel version of \textlcsc{SNAPDRAGONS} \citep{HKG15} denoted \Hmocks, and another with the code presented in \citet{LWC15} denoted \Dmocks. These catalogues contain the true and observed phase space coordinates of stars, their \emph{Gaia} DR2 errors, magnitudes in several passbands, metallicities, ages, masses, and stellar parameters. We show that a powerful use of the mock catalogues is to compare them with the intrinsic simulation data from which they were generated in order to acquire predictions of how accurately physical properties are reproduced, and to determine which kind of data should be studied from the \emph{Gaia} survey to target specific questions. We focus on two practical applications: the structure of the young stellar disc and kinematics of the stellar halo. In particular, we show that, in contrast to typical disc setups in many idealised $N$-body simulations, the \textlcsc{AURIGA} simulations predict that young stars ($\sim$few hundred Myr old) make up flared distributions (increasing scale height with increasing radius), which are well traced by B- and A-dwarf stars. We also show that the systemic rotation of the stellar halo can be accurately inferred from \emph{Gaia} data. Finally, we discuss the limitations of our methods and provide information on how the community can access the mock data.

\section{Magneto-hydrodynamical Simulations}
\label{sec2}

\begin{table*}
\centering
\caption{Table of properties of each simulation. The columns are: 1) halo number; 2) virial mass 3) virial radius; 4) stellar mass within the virial radius; 5) stellar disc mass calculated as $2\pi \Sigma _0 R_d^2$, where $\Sigma _0$ and $R_d$ are the parameters retrieved from a bulge-disc surface density decomposition performed in the same way as in \citet{GGM17} for the mass within 1 kpc of the disc midplane; 6) stellar disc scale length; 7) circular rotation velocity at a radius of 8 kpc, calculated as $V_c = \sqrt{GM(<R=8\, {\rm kpc}) / 8\,{\rm kpc}}$; 8) azimuthally averaged stellar surface density within 1 kpc of the midplane at $R=8$ kpc; 9) thin and thick (bracketed values) disc scale heights of a double $\rm sech^2$  decomposition of the vertical density distribution in a 1 kpc-wide annulus centred at $R=8$ kpc (see Fig.~\ref{figsch}); 10) vertical velocity dispersion of stars within 1 kpc of the disc midplane at $R=8$ kpc. The last row provides current estimates of all of these quantities for the Milky Way. All values are taken directly from \citet{BHG16}. $^{\dagger}$The mean of values for $R_{200}$ provided in Table. 8 of that paper, the standard deviation of which is $28.6$. $^{\ddagger}$Observationally derived vertical scale height and velocity dispersion of the old thin disc and thick disc (bracketed values) at the solar neighbourhood.}
\begin{tabular}{c c c c c c c c c c}
\hline
Run & $\frac{M_{\rm vir}}{[10^{12} \rm M_{\odot}]}$ & $\frac{R_{\rm vir}}{[\rm kpc]}$ & $\frac{M_{*}}{[10^{10} \rm M_{\odot}]}$ & $\frac{M_{*,\rm d}}{[10^{10}\rm M_{\odot}]}$ & $\frac{R_{\rm d}}{[\rm kpc]}$ & $\frac{V_{\rm c}\,(R_{\odot})}{[\rm km\, s^{-1}]}$ & $\frac{\Sigma \,(R_{\odot})}{[\rm M_{\odot} \, pc^{-2}]}$ & $\frac{h_z\,(R_{\odot})}{[\rm pc]}$ & $\frac{\sigma_{z}\,(R_{\odot})} {[\rm km\, s^{-1}]}$  \\
\hline                                                                     
Au 6   & 1.01 & 211.8 & 6.1 & 2.6 & 3.3 & 224.7 & 33.2 & 339 (1139) & 39.8 \\
Au 16  & 1.50 & 241.5 & 7.9 & 3.7 & 6.0 & 217.5 & 44.5 & 303 (1130) & 40.2 \\
Au 21  & 1.42 & 236.7 & 8.2 & 3.8 & 3.3 & 231.7 & 51.8 & 430 (1363) & 44.0 \\
Au 23  & 1.50 & 241.5 & 8.3 & 4.0 & 5.3 & 240.0 & 52.5 & 339 (1260) & 42.0 \\
Au 24  & 1.47 & 239.6 & 7.8 & 2.8 & 6.1 & 219.2 & 31.5 & 330 (1436) & 42.4 \\
Au 27  & 1.70 & 251.4 & 9.5 & 5.0 & 3.2 & 254.5 & 71.1 & 302 (1103) & 42.1 \\
\hline
MW & $1.3 \pm 0.3$ & $^{\dagger}220.7$ & $6\pm 1$ & $4\pm 1$ & $2.6\pm 0.5$ & $238 \pm 15$ & $33.3 \pm 3$ & $^{\ddagger}300\pm 50$ ($900\pm 180$) & $^{\ddagger}25\pm 5$ ($50\pm 5$) \\
\hline
\end{tabular}
\label{table1}
\end{table*}

\begin{figure*}
\includegraphics[scale=0.3,trim={0 0 0 0},clip]{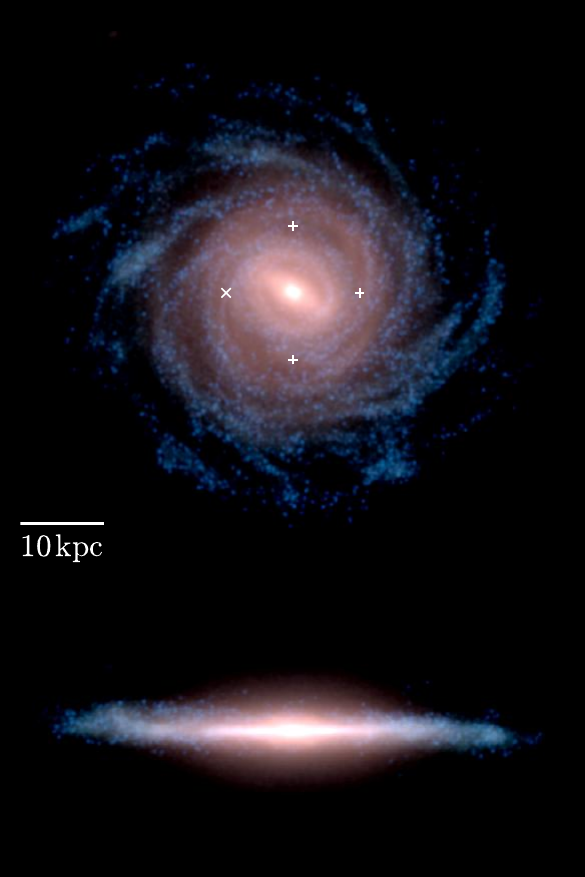}\hspace{-0.1cm}
\includegraphics[scale=0.3,trim={0 0 0 0},clip]{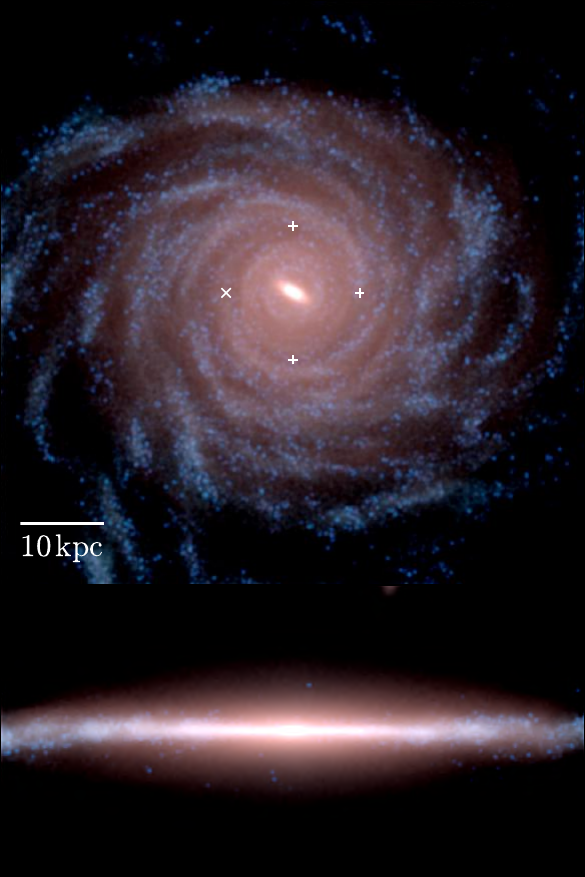}\hspace{-0.1cm}
\includegraphics[scale=0.3,trim={0 0 0 0},clip]{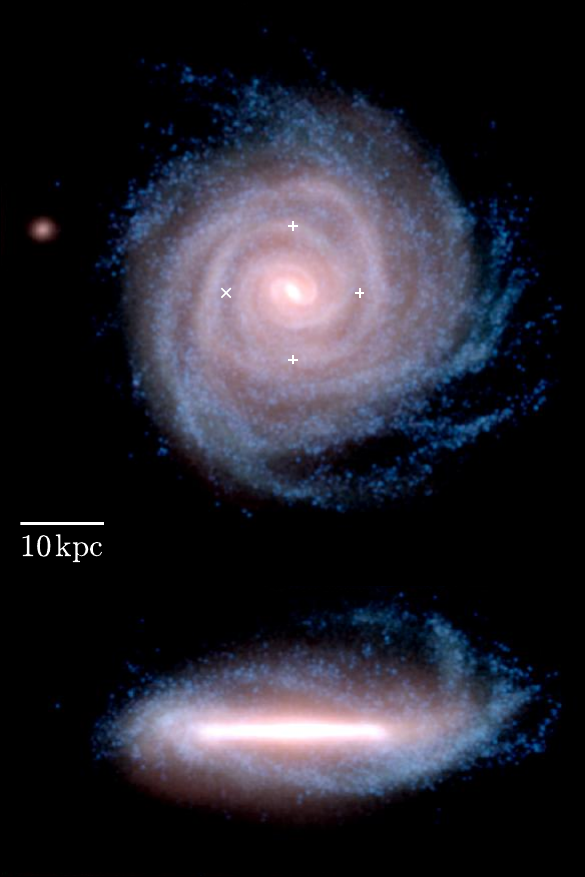}\hspace{-0.1cm}
\includegraphics[scale=0.3,trim={0 0 0 0},clip]{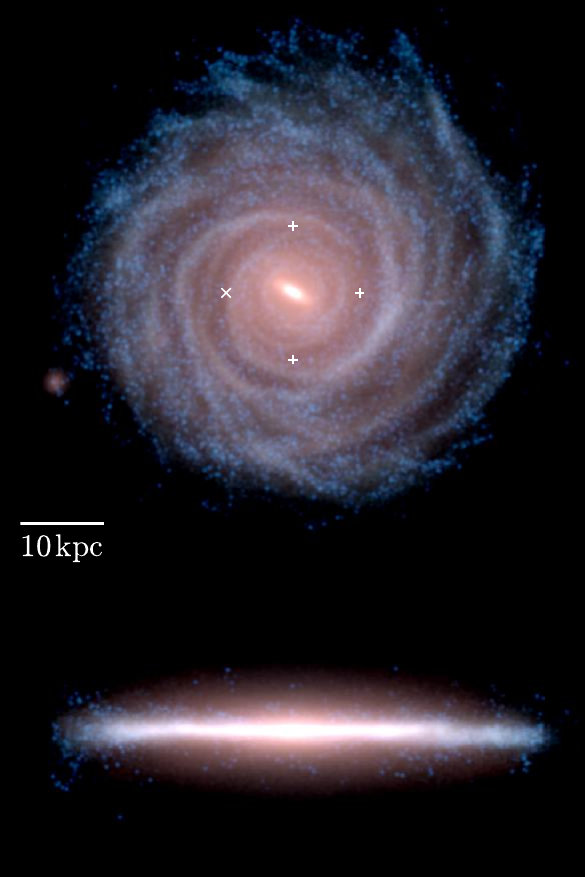}\hspace{-0.1cm}
\includegraphics[scale=0.3,trim={0 0 0 0},clip]{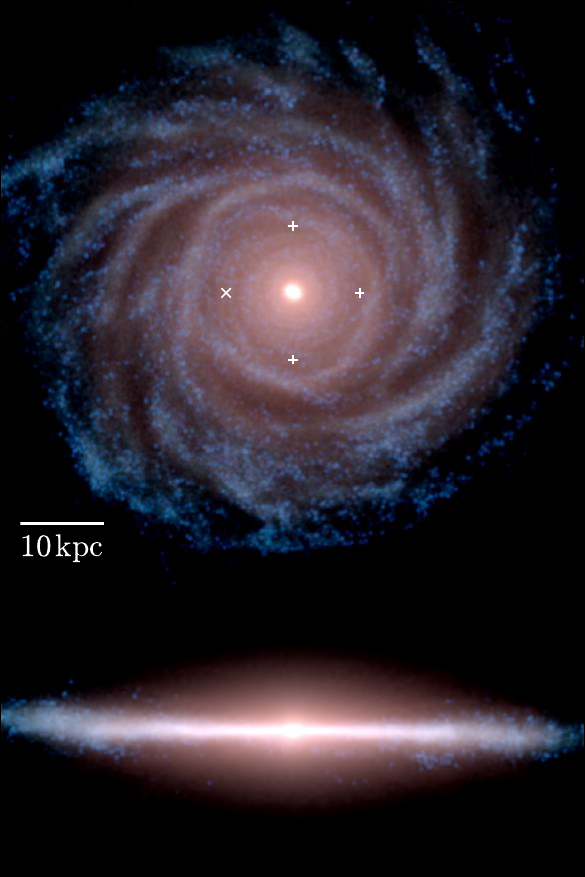}\hspace{-0.1cm}
\includegraphics[scale=0.3,trim={0 0 0 0},clip]{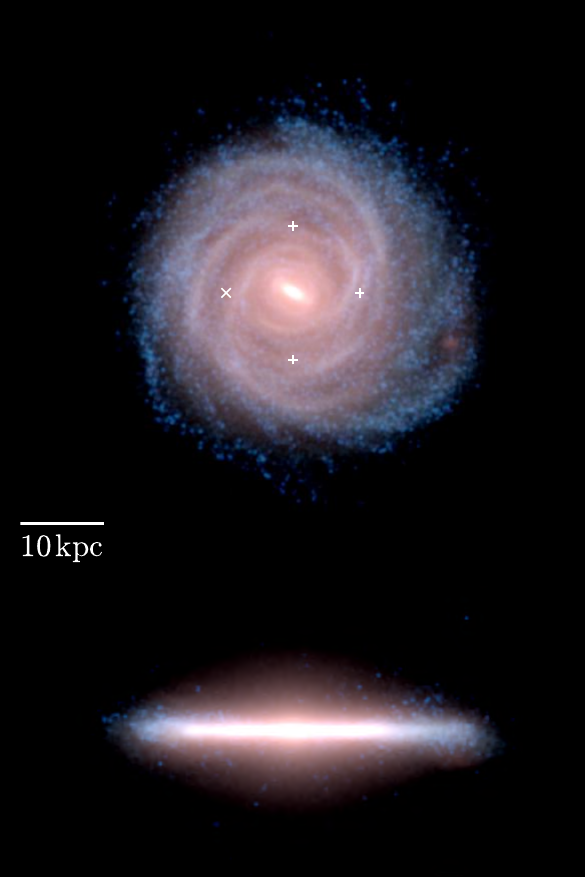}
\caption{Face-on and edge-on projected stellar densities at $z=0$ for
  the six high-resolution simulations from which we construct mock
  catalogues. The images are a projection of the $K$-, $B$- and
  $U$-band luminosity of stars, shown by the red, green and blue
  colour channels, in logarithmic intervals, respectively. Younger
  (older) star particles are therefore represented by bluer (redder)
  colours. The box side-length is 70 kpc in each panel. The cross
    in each panel (leftmost white symbol) indicates the default Solar position, whereas the
    plusses indicate the other choices. Movies and images are
  available at
  \href{http://auriga.h-its.org}{\url{http://auriga.h-its.org}}.} 
\label{figau}
\end{figure*}

\begin{figure*}
\includegraphics[scale=0.9,trim={2.1cm 0 0 0},clip]{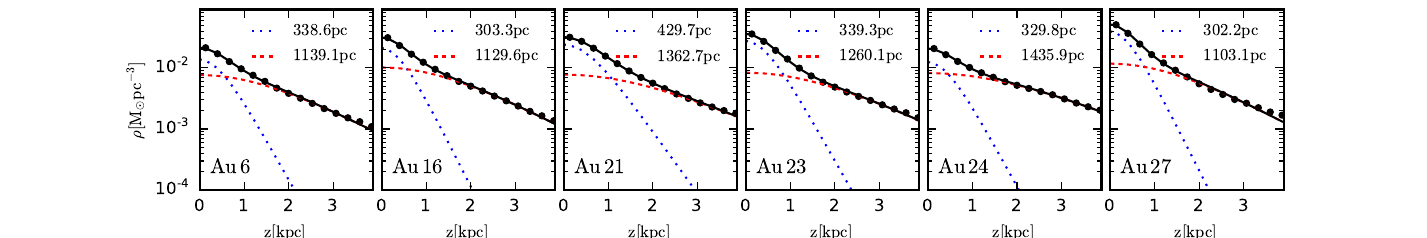}
\caption{Profiles of the stellar vertical density distribution in
    a 1 kpc-wide annulus centred at $R=8$ kpc, for each simulation. A
    double $\rm sech^2$ profile is fitted (black curves) to the raw
    density distribution (black circles). The profile is composed of a
    thin (blue dotted curves) and a thick (red dashed curves) disc,
    the scale heights of which are denoted in each panel. In each
    case, values similar to the Milky Way values are obtained (see
    Table. 1).}
\label{figsch}
\end{figure*}

The \textlcsc{AURIGA} simulations \citep{GGM17} are a suite of cosmological zoom simulations of haloes in the virial mass\footnote{Defined to be the mass inside a sphere in which the mean matter density is 200 times the critical density, $\rho _{\rm crit} = 3H^2(z)/(8 \pi G)$.} range $10^{12}$ - $2\times10^{12}$ $\rm M_{\odot}$. The haloes were identified as isolated haloes\footnote{The centre of a target halo must be located outside of $9$ times the $R_{200}$ of any other halo that has a mass greater than $3\%$ of the target halo mass.} from the redshift $z=0$ snapshot of a parent dark matter only simulation with a comoving side length of 100 cMpc from the EAGLE project (L100N1504) introduced in \citet{SCB15}. Initial conditions for the zoom re-simulations of the selected haloes were created at $z=127$, using the procedure outlined in \citet{J10} and assuming the \citet{PC13} cosmological parameters: $\Omega _m = 0.307$, $\Omega _b = 0.048$, $\Omega _{\Lambda} = 0.693$ and a Hubble constant of $H_0 = 100 h$ km s$^{-1}$ Mpc$^{-1}$, where $h = 0.6777$. The halos are then re-simulated with full baryonic physics with higher resolution around the main halo.

The simulations were performed with the magneto-hydrodynamic code
\textlcsc{AREPO} \citep{Sp10}, and a comprehensive galaxy formation
model \citep[see][for more details]{VGS13,MPS14,GGM17}. This
  model includes atomic and metal line cooling \citep{VGS13} and a spatially uniform UV background \citep{FG09}, which fully reionizes hydrogen at
  redshift 6. A subgrid model for the interstellar medium and star
  formation \citep{SH03} is employed. Stellar evolution is treated
  self-consistently and includes metal enrichment from core collapse
  supernovae, thermonuclear supernovae, and AGB stars
  \citep{VGS13}. Feedback from core collapse supernovae is taken into
  account through a non-local effective wind model that isotropically
  carries thermal and kinetic energy in equal partition away from the
  star-forming ISM. The winds are hydrodynamically decoupled from the
  gas until they encounter gas with a density below 10 per cent of the
  star formation threshold, at which time they are recoupled to the
  gas. As a result, the winds deposit mass, metals and energy, and
  impart momentum predominantly to gas that surrounds the star-forming
  regions. The metal content of the winds is equal to $1-\eta$ times
  the total mass of metals of the star-forming gas from which they are
  launched, where $\eta = 0.6$. The rate at which winds are launched
  is determined by the mass loading factor, which depends on the
  energy available for supernovae (given by the star formation rate)
  and the wind velocity, which we set equal to 3.46 times the local 1D
  dark matter velocity dispersion \citep{OFJ10}. 

Star particles are assumed to be simple stellar populations (SSPs), and are assigned broad band luminosities based on the catalogues of \citet{BC03}. Stellar mass loss and metal enrichment are modelled by calculating at each time step the mass (and metal content thereof) moving off the main sequence for each star particle according to a Chabrier \citep{C03} Initial Mass Function, which is then distributed isotropically into surrounding gas cells. Lower and upper mass limits of 0.1 and 100 $\rm{M}_\odot$, respectively, are set for the integration limits. The mass and metals are then distributed among nearby gas cells with a top-hat kernel. We track a total of 9 elements: H, He, C, O, N, Ne, Mg, Si and Fe. 

The model further includes the seeding, growth and feedback from
  supermassive black holes \citep{SMH05}. The gas accretion rate is
  given by the Bondi-Hoyle-Lyttleton model \citep[][BHL
  hereafter]{BH44,B52}, 
  and a term designed to balance the energy lost from the intracluster
  medium (ICM) of the halo in the form of X-ray emission is  based on the
  model of \citet[][NF hereafter]{NF00}. The BHL accretion rate gives
  rise to thermal energy feedback injected into surrounding gas cells,
  whereas the NF term provides thermal energy required to inflate
  small bubbles of hot gas in the halo in a smooth fashion. 

Magnetic fields are included in the limit of ideal
  magnetohydrodynamics (MHD), which is described in detail in
  \citet{PakS13} and \citet{PGG17}. A uniform magnetic seed field with
  comoving strength, $10^{-14}$ G, is set at $z = 127$ (equal to a
  physical strength of $2 \times 10^{-4}\mu$ G) oriented along the
  z-coordinate of the simulation cube. We note that although this seed
  field strength is many orders of magnitudes larger than plausible
  values for a cosmological seed field from inflation or fields seeded
  by Biermann batteries \citep{KZ08}, the information about the
  initial configuration and strength of the magnetic field is quickly
  erased by an exponential dynamo in collapsed halos
  \citep{PMS14,MVM15}. The initial strength is sufficiently small to
  be dynamically irrelevant outside collapsed halos \citep{MV16}.  
 
In this paper, we focus on the highest resolution simulations of the
\textlcsc{AURIGA} suite, which correspond to the ``level 3''
resolution described in \citet{GGM17}. The galaxies were selected
  to have large discs (Au 16, Au 24), to be close analogues of the
  Milky Way as measured, for example, by stellar mass, star formation
  rate and morphology (Au 6), or for interesting satellite
  interactions (Au 21, Au 23, Au 27). The typical dark matter
particle mass is $\sim 4 \times 10^{4}$ $\rm M_{\odot}$, and the
baryonic mass resolution is $\sim 5 \times 10^{3}$ $\rm
M_{\odot}$. The physical softening of collisionless particles
increases with time up to a maximum physical softening length of 185
pc, which is reached at redshift 1. The physical softening value for
the gas cells is scaled by the gas cell radius (assuming a spherical
cell shape given the volume), with a minimum softening set to that of
the collisionless particles.

Final face-on and edge-on stellar luminosity images for these systems
are shown in Fig.~\ref{figau}. We list some relevant properties of the
simulations in Table.~\ref{table1}. The disc scale lengths, derived
from fits to the surface density distribution of stars $\leq 1$ kpc of
the midplane, range from 3.2 kpc to 6.1 kpc, and implied stellar disc
masses from 2.6 $\times 10^{10}\rm M_{\odot}$ to 5
$\times 10^{10}\rm M_{\odot}$, which are similar to current estimates
for the Milky Way \citep{BHG16}. We remark that each of the
  simulated discs can be decomposed into a thick and thin disc at
  $R\sim 8$ kpc (see Fig.~\ref{figsch} and Table.~1) with scale height
  values similar to those of the Milky Way. The simulated vertical
  velocity dispersion is calculated from all stars within 1 kpc of the
  disc midplane at $R\sim 8$ kpc, and falls between the thin and thick
  disc values derived observationally.
  \citep[e.g.][]{Mc11}. The ability of these simulations to produce
coherent, radially extended discs with barred and spiral
structure and stellar haloes from a self-consistent cosmological
galaxy formation model from $\Lambda$CDM initial conditions makes
these simulations powerful predictors for the formation of galaxies
like the Milky Way. In the next section, we describe how we generate
the mock \emph{Gaia} catalogues from the simulations.

\section{Mock stellar catalogues}
\label{sec3}

The first step to create a mock stellar catalogue is to choose the
position and velocity of the Sun. For each simulation, we define four
choices for the solar position: all adopt a radius and height above
the midplane (defined at redshift 0) of $(R_{\odot},Z_{\odot}) =
(8,0.02)$ kpc, and are spread at equidistant azimuthal angles relative
to our default reference angle, which is chosen to be 30 degrees
behind{\footnote{\emph{Behind} means an angle measured from the bar
    major axis in the direction opposite to that of the rotation of
    the Galactic disc. We note that an effectively random
      azimuthal position is chosen for Au 24, which does not have a
      bar.}} the major axis of the bar (Bland-Hawthorn \& Gerhard
2016). The bar major axis is calculated from the $m=2$ Fourier mode of
the central 5 kpc stellar distribution \citep[see][for details on how
to extract angles from modes]{GKC13}. We then rotate the disc such
that the solar position is placed at the Galactocentric Cartesian
coordinate $(X,Y,Z)=(-R_{\odot}, 0, Z_{\odot})$. We set the local
standard of rest equal to the spherically averaged circular velocity
at the solar radius, and set the Solar motion velocity to $(U_{\odot},
V_{\odot}, W_{\odot}) = (11.1, 12.24, 7.25)$ $\rm km \, s^{-1}$
\citep{SBD10} relative to the local standard of rest. After setting
the solar position and velocities, we transform our coordinate system
to heliocentric equatorial coordinates following the matrix
transformation described in Section 3 of \cite{HK14}, and we retain
this coordinate system in the mock catalogue output. 

For each of the four Solar positions, we generate two sets of mock catalogues: one set is generated by a parallelised version of \textlcsc{SNAPDRAGONS}\footnote{Serial version available at \href{https://github.com/JASHunt/Snapdragons}{\url{https://github.com/JASHunt/Snapdragons}}} \citep{HKG15} (\Hmocks{}); the other set is generated using the method presented in \citet{LWC15} (\Dmocks{}), who produced SDSS mocks based on the the \citet{CCF10} particle tagging technique applied to the \textsc{aquarius} simulations \citep{SWV08}. \citet{Mateu:2017aa} added \gaia{} observables to the \citeauthor{LWC15} mocks to make predictions for the detection of tidal streams in \gaia{} data using great-circle methods.

Both methods assume that each simulation star particle is a Simple Stellar Population (SSP) that can be transformed into individual stars by sampling from a theoretical isochrone matching the particle's age and metallicity. They compute observable properties of stars and their associated errors in the same way, and apply identical selection functions. The methods differ in how the stars are distributed in phase space and their choice of stellar evolution models. The step-by-step procedure for generating each set of catalogues is as follows:

\paragraph*{\Hmocks}
\begin{enumerate}
	\item apply a stellar population synthesis model to each star particle; \smallskip
    \item add  dust extinction;\smallskip
    \item apply the observational selection based on a magnitude cut;\smallskip
    \item convolve observable properties with \emph{Gaia} DR2 errors and displace stellar coordinates.
\end{enumerate}

\paragraph*{\Dmocks}
\begin{enumerate}
	\item apply a stellar population synthesis model to each star particle; \smallskip
    \item add  dust extinction;\smallskip
    \item distribute individual stars over the approximate phase space volume of the parent star particle;\smallskip
    \item apply the observational selection based on a magnitude cut;\smallskip
    \item convolve observable properties with \emph{Gaia} DR2 errors and displace stellar coordinates.
\end{enumerate}

We note that the \Hmocks{} displace stars from their parent particles (true coordinates) to their observed coordinates by random sampling the DR2 error distributions for astrometry and radial velocity of the mock star. However, the \Dmocks{} distribute stars over a 6D kernel approximating the phase-space volume of their parent particle, which become the true coordinates, and are afterwards displaced to their observed coordinates by error sampling in the same way as the \Hmocks. In addition, we generate a version of the \Dmocks{} without extinction by omitting step (ii), which we denote as \Dmocksnoex. We discuss the advantages and disadvantages of this choice below, where we describe each stage in detail.

\subsection{Stellar Population Synthesis}
\label{sec:popsynth}

The basic premise of the population synthesis calculation in both the \Hmocks{} and \Dmocks{} is that each simulation star particle corresponds to an SSP with an evolutionary state defined by a single metallicity and age, and a total number of stars proportional to its mass. The present day mass distribution of individual stars in the SSP is determined by the convolution of an assumed IMF by a model of stellar evolution (encapsulated in a set of pre-computed isochrones), which takes into account processes such as the death of massive stars and mass loss from those that survive.

For the \Hmocks, although the simulations use a Chabrier IMF, \sc{snapdragons }\rm only contains implementations of the Salpeter \citep{S55} and Kroupa \citep{K01} IMFs. Thus, we use a Kroupa IMF to sample the distribution of present-day stellar masses for each SSP which is the closer approximation of the Chabrier IMF used in the \textlcsc{AURIGA} simulations. We set the minimum allowed initial stellar mass to be 0.1 $\rm M_{\odot}$ (as for the \textlcsc{AURIGA} simulations). For a given SSP, we set the lower mass limit to be the lowest present day stellar mass that would be visible at our limiting magnitude (see below), and the upper stellar mass limit to be the maximum stellar mass which would still be present at the age of our model particle. We then integrate the IMF over the desired mass range to determine the number of stars which would be visible within this mass range, $N_{\mathrm{s}}$, and randomly sample the IMF $N_{\mathrm{s}}$ times. Note that while we do not generate any stars below the visible limit, we do account for their mass. The process is discussed in more detail in \citet{HKG15}.

The procedure described above is similar for the \Dmocks, which use a Chabrier IMF. To sample the SSP, we choose small intervals of initial mass in the range\footnote{We note that the lower mass limit of $0.08$ is lower than the limit of $0.1$ adopted by the \textlcsc{AURIGA} simulations, however, $\rm M7V$-$\rm M8V$ stars of this mass have an absolute $V-$band magnitude of $\sim 18$ (fainter than our $V<16$ allsky sample) and an apparent magnitude fainter than $V=20$ at distances farther than 25 pc from the Sun (with no extinction). These extremely faint stars will therefore not be observed for the vast majority of applications. The upper mass limit of $120~\rm{M}_\odot$ is higher than the $100~\rm{M}_\odot$ assumed in \textlcsc{AURIGA}; however such massive stars are extremely rare, therefore we do not expect them to bias any results.} $0.08$ to $120~\rm{M}_\odot$. Given the total initial mass of the SSP, we calculate the expected number of stars in each interval. Finally, the actual number of stars in each mass interval is randomly generated from a Poisson distribution with the corresponding expectation value. 

Once we have sampled the stellar mass distribution for a given star particle, we are in a position to assign stellar parameters such as temperature, magnitudes, and colours to each synthetic star. For the \Dmocks, we use the \textsc{parsec} isochrones \citep{Bressan2012,Tang2014,Chen2014,Chen2015}. These represent up-to-date stellar models that span a wide range of metallicities and ages, and have magnitudes in multiple bands, including the \gaia{} ones. We downloaded isochrone tables from the \texttt{CMD v3.0} web interface\footnote{\href{http://stev.oapd.inaf.it/cgi-bin/cmd_3.0}{\url{http://stev.oapd.inaf.it/cgi-bin/cmd_3.0}}} using the default options. We sample a grid of isochrones spanning the age range $6.63 \leq \log(t/\rm{yr}) \leq 10.13$, with a step size, $\Delta \log (t/\rm{yr}) = 0.0125$, and the metallicity range $0.0001 \leq Z \leq 0.06$. Because interpolating between precomputed isochrones is nontrivial, we identify the isochrone with the closest value in age and metallicity for each star particle. Any particles that lie outside the range of the age/metallicity grid are also matched to the nearest isochrone.

For the \Hmocks, we use the same procedure as described above, but use an earlier version of the PARSEC isochrones \citep{MGBGSG08}, which are currently used in the \textlcsc{SNAPDRAGONS} code. This set of isochrones uses a slightly different range of ages and metallicities for the grid compared to those used for the \Dmocks{}: $6.6 \leq \log(t/\rm{yr}) \leq 10.22$, with a step size, $\Delta \log (t/\rm{yr}) = 0.02$ and $0.0001 \leq Z \leq 0.03$. We do not expect that the properties of most stellar populations in our catalogues will be significantly affected by the differences between these two sets of isochrones.

\subsection{Dust Extinction}
\label{sec:dext}

Dust extinction can be problematic for Galactic optical surveys, such as \emph{Gaia}, mainly because of the poorly understood three-dimensional distribution of dust in the Milky Way. As an approximation, the \Hmocks{} use the extinction maps used in \textlcsc{Galaxia} \citep{SBJ11}, based on the method presented in \citet{BKF10} to derive a 3D polar logarithmic grid of the dust extinction generated from the 2D dust maps of \citet{SFD98} and the assumption of a uniform distribution of dust along a given line of sight. From these maps, we calculate a magnitude extinction for each magnitude band and, given the distance modulus for the original star particle, we determine the apparent magnitude in each band. 

We note that the alternative philosophy of modelling dust directly from the gas  and dust distribution in the simulations will make the dust map more consistent with large-scale features of the \textlcsc{AURIGA} galaxies (such as spiral arms). However, going beyond uncertain, simplistic dust models based solely on the metallicity of simulation gas cells is far from straightforward \citep[e.g.][]{Trayford2017}. On the other hand, the use of a dust map based on the Milky Way results in one fewer discrepancy between the mock catalogues and observations that use the same dust maps; this may facilitate their inter-comparison because the selection function will be more consistent with \emph{Gaia}.

The \Dmocksnoex{} do not include dust extinction, and hence the user is free to adjust magnitudes for extinction themselves, if required. We note also that dust extinction is less important for stellar halo studies, which typically exclude high extinction regions in the Galactic mid-plane.

\subsection{Phase space sampling}
\label{sec:phase_space_sampling}

{\it This step is applied only to the \Dmocks.} Once we have generated a catalogue of stars, the \Dmocks{} method assigns distinct positions in configuration and velocity space to each of them. The intention of this step, which can be thought of as a form of smoothing, is to avoid discrete `clumps' of stars at the coordinates of the parent particles. We follow the implementation of \citet{LWC15}, which is similar to that introduced by the \textlcsc{Galaxia} code \citep{SBJ11}. For every simulation particle we construct a six-dimensional hyper-ellipsoidal `smoothing kernel' that approximates the volume of phase space the particle represents. We distribute the stars associated with particles into these 6D kernels as described below. In this way, we approximately preserve coherent phase space structures in the original simulation, such as tidal streams (e.g. in configuration space, this approach ensures stars are displaced more along such streams than they are perpendicular to them). It is important to note that, although the resulting distribution of stars represents a denser sampling of phase space, it is essentially an interpolation (and extrapolation, around the edges of the phase space of the simulation). It does not add any (physical) dynamical information or increase the resolution beyond that of the parent simulation.

The phase-space volume associated with each star particle is estimated using the \enbid{} code of \citet{Sharma2006}. This code numerically estimates the 6D phase space density around each particle by using an entropy based criterion to partition the set of particles into a binary tree, without the need to specify a metric relating configuration and velocity space. The resulting estimate of the phase-space volume of each leaf node can be noisy due to Poisson sampling, so we further apply an anisotropic smoothing kernel. We use the nearest 64 neighbours to locally determine the principal directions and to locally rescale the phase space. In this rotated and rescaled phase space, we define the phase space volume, $V_{6D}$, of each star particle as $1/40$ of the hypersphere which encloses the nearest 40 neighbours. The actual phase-space sampling kernel is a 6D isotropic Gaussian with zero mean and dispersion, $\sigma^2=\gamma R_{6D}^{2}$, where $\gamma=1/48$ and $R_{6D}$ is the radius of the hypersphere with volume, $V_{6D}$. To avoid extreme outliers in the Gaussian tails of these kernels, we truncate the kernels at $5\sigma$. We draw coordinates randomly from the kernel defined by each parent star particle for each star it generates. Each randomly generated point is then transformed back from this rotated and rescaled phase space into the Cartesian configuration and velocity space of the original simulation. We call these new coordinates the ``true'' coordinates. This definition differs from that in the \Hmocks, in which the ``true'' positions correspond to those of the parent star particle. See \citet{LWC15} for a more detailed description and several tests of the phase space sampling method. 

To avoid unnecessary over-smoothing due to `cross-talk' between different phase-space structures, we partition the stellar particles into sets according to their progenitor galaxy, and calculate the scale of the phase space kernels for a given particle using only neighbours from the same set. For this purpose we use the \textlcsc{AURIGA} merger trees built from \textsc{subfind} groups \citep{GGM17}. We trace back each stellar particle to the first snapshot in which it belonged to the same FOF halo as the main progenitor of the Milky Way halo analogue. Particles which did not form `in situ' in the central galaxy are grouped according to their subfind group membership at the snapshot immediately prior to this (i.e. just before their first infall into the main progenitor halo). We assign all particles which did form in the central galaxy to a single group (we discus a potential limitation of this implementation in Sec.~\ref{sec:limitations}). Again, further details are given in \citet{LWC15}.

\begin{figure*}
	\includegraphics[scale=0.48,trim={0 3.5cm 0 1.cm},clip]{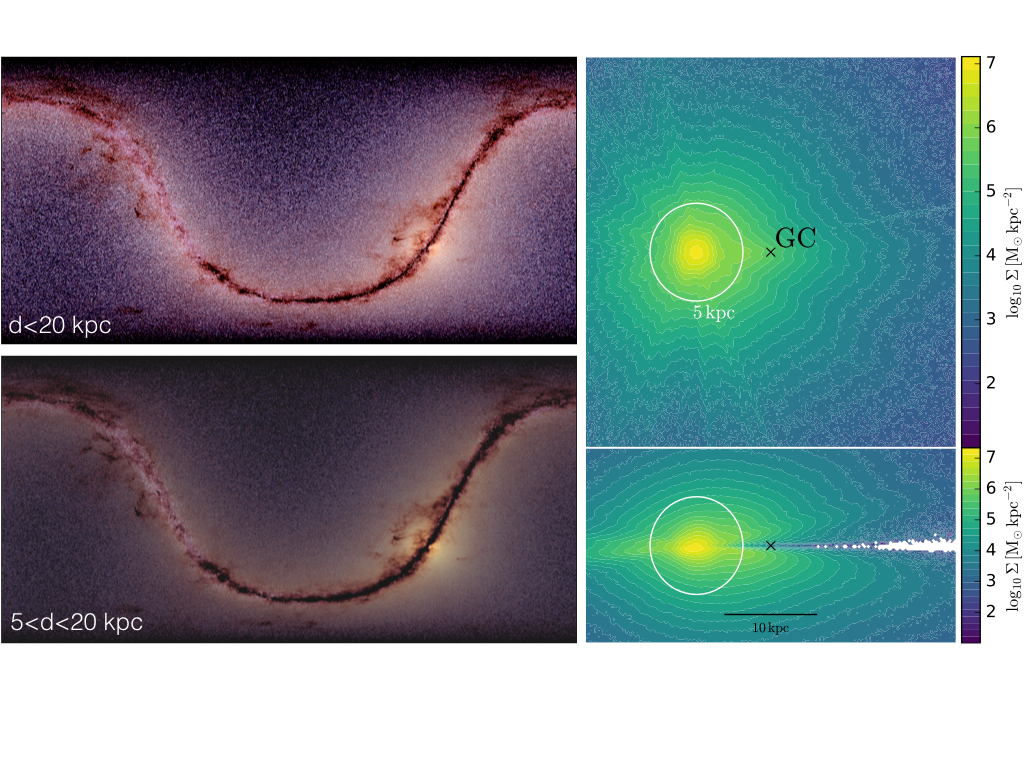}
	\caption{\emph{Left:} Three-colour all-sky maps in heliocentric equatorial coordinates of the default \textsc{hits-mock} for Au 24. These maps are constructed from mapping the $K$-, $G$- and $U$-band apparent magnitudes to the red, green and blue colour channels of the composite image. The $x$ and $y$ axes represent right ascension (RA) and declination (DEC), respectively. The upper image shows the stellar light distribution for all stars up to a 20 kpc heliocentric distance, $\rm d$, whereas the lower image shows the map for stars between 5 and 20 kpc heliocentric distance. \emph{Right:} Contour maps show the projected face-on (top panel) and edge-on (bottom panel) stellar mass surface density, respectively, with annotations for the Galactic centre and  5 kpc heliocentric distance, to guide the eye. }
	\label{fig1}
\end{figure*}

\subsection{Mock survey selection function}

In order to limit the size of our mock catalogues to the order of $\sim10^8$ stars instead of $\gtrsim10^9$ stars, we provide a full sky catalogue only for stars with $V<16$. Most stellar halo stars are fainter than this, so to have a large sample of stars for stellar halo science we supplement this bright star catalogue by including stars with $16 < V < 20$ for Galactic latitudes $|b|>20$ degrees. These selection cuts are applied to both the \Hmocks{} and the \Dmocks. 

We note that in the \Hmocks, faint stars are randomly sampled at a rate of $20\%$ in order to reduce the output size. However, this does not bias data trends aside from the number of stars available in the magnitude range $16 < V < 20$.

\subsection{\emph{Gaia} DR2 errors}

In this subsection, we describe how we add \emph{Gaia} DR2 errors to the catalogues, which is the same for both the \Hmocks{}  and \Dmocks. We convolve the parameters of the selected stars with \emph{Gaia}-like errors as a function of magnitude and colour in the Johnson-Cousins $V$ and $I_{c}$ bands following \citet{JGC10},
\begin{equation}
\begin{aligned}
G = V - 0.0257 - 0.0924(V-I_c) - 0.1623(V-I_c)^2 \\+ 0.009(V-I_c)^3.
\end{aligned}
\end{equation}
We use the post-launch error estimates approximated from the estimates in pre-launch provided through the Gaia Challenge collaboration performance \citep[][]{R-G+15}, which include all known instrumental effects such as stray light levels and residual calibration errors. A simple performance model that takes into account the wavelength dependence of the point spread function and reproduces the end-of-mission parallax standard error estimates, is
\begin{equation}
\begin{aligned}
\sigma _{\pi _{\mathrm{final}}} [\mu \rm as] = (-1.631 + 680.766  z + 32.732  z^2)^{0.5} \\ \times [0.986 + (1 - 0.986)  (V-I_c)],
\end{aligned}
\end{equation}
where
\begin{equation}
z = {\rm max}\left(10^{0.4  (12.09 - 15)}, 10^{0.4  (G - 15)}\right),
\end{equation}
and $6\le G\le 20$ denotes the range in broad-band, white-light, \emph{Gaia} magnitudes. This relation reflects the magnitude-dependent errors for stars observed by \emph{Gaia}. Stars in the range $6\le G\le 12$ will have shorter integration times in order to avoid CCD saturation, and are assigned a constant $\sigma _{\pi} = 7$ $\mu$as error by the above relation. 

The basic mission results improve with increasing mission time, $t$, as $t^{-0.5}$ for the positions, parallaxes, photometry and radial velocities, and $t^{-1.5}$ for the proper motions\footnote{\href{http://www.astro.lu.se/gaia2017/slides/Brown.pdf}{\url{http://www.astro.lu.se/gaia2017/slides/Brown.pdf}}}. Given that these errors are end-of-mission estimates, we adopt the following simple scaling to provide the expected parallax-standard error for DR2: 
\begin{equation}
\sigma _{\pi} = L \sigma _{\pi _{\rm final}},
\end{equation}
where $L=(60/22)^{1/2}$, which corresponds to the square root of the DR2 mission time divided by the total 5 year mission time. The right ascension, declination and proper motions are all scaled with this factor as well.

The errors in position on the sky ($\alpha$, $\delta$) and proper motions ($\mu _{\alpha}$, $\mu _{\delta}$) scale with the ecliptic longitude averaged error of the sky-varying factors derived from scanning law simulations, the values of which are listed on the \emph{Gaia} performance website\footnote{\href{https://www.cosmos.esa.int/web/gaia/science-performance}{\url{https://www.cosmos.esa.int/web/gaia/science-performance}}}. 

DR2 will provide radial velocities for only a very small subset of stars near the Sun with spectral type later than $\rm F$. However, the selection function and error function is non-trivial, involving, for example, the number of visits, binarity and temperature. Thus, we provide estimates of the radial velocity error for all generated stars, using the end of mission \emph{Gaia} error which adopts the simple performance model, 
\begin{equation}
\sigma _{v_r} = 1 + b e^{a(V-12.7)},
\end{equation}
where $a$ and $b$ are constants that depend on the spectral type of the star. We caution the reader that the radial velocities are both more plentiful and more accurate than the expected DR2 radial velocities.

In addition to astrometric errors, we calculate the red and blue broadband \emph{Gaia} magnitudes, $G_{\rm RP}$ and $G_{\rm BP}$, and errors for all \emph{Gaia} photometric bands, according to the single-field-of-view-transit standard error on the \emph{Gaia} science performance website, modified to include the DR2 mission time scaling and $20\%$ calibration errors:
\begin{align}
\sigma _G = & \,\,5\frac{1.2\times10^{-3} L}{\sqrt{70}} \nonumber \\ & \left(0.04895 z^2 + 1.8633 z + 0.0001985\right)^{1/2},
\label{Gsig}
\end{align}
and
\begin{align}
\sigma _{G_{\rm RP/BP}} = & \,\,5\frac{1\times 10^{-3} L}{\sqrt{70}} \nonumber \\ & \left(10^{a_{\rm BP/RP}} z^2 + 10^{b_{\rm BP/RP}} z + 10^{c_{\rm BP/RP}}\right)^{1/2},
\label{GRPsig}
\end{align}
where $a_{\rm BP/RP}$, $b_{\rm BP/RP}$ and $c_{\rm BP/RP}$ are listed on the \emph{Gaia} science performance website. We note that the factor of 5 in the pre-factor of equations~(\ref{Gsig}) and (\ref{GRPsig}) is required to scale the photometric errors to match the $\sim$ millimag accuracy at the bright end ($G < 13$ mag) and the 20 millimag and 200 millimag accuracy at the faint end for $G$ and $G_{\rm RP/BP}$, respectively, that are quoted on the \emph{Gaia} DR2 website.

We provide error estimates for atmospheric parameters based on the results of \citet{LBJ12}, who inferred the expected performance of stellar parametrisation from various fitting methods applied to synthetic spectra. Specifically, a second order polynomial in $G$ has been fitted to the mean averaged residual of effective temperature and surface gravity inferred from the Bayesian method Aeneas \citep{BJ11}. 

For both the \Hmocks{} and the \Dmocks, we randomly sample these standard errors for each generated mock star (that satisfies our magnitude cut) to displace the measured parallax, proper motions and radial velocity of each synthetic star from that of its parent particle. This ensures that, for the reasons discussed in Sec.~\ref{sec:phase_space_sampling}, the position and velocity coordinates of each star are distinct from those of their parent star particle in the case of the \Hmocks. The standard errors for the \emph{Gaia} photometric bands (equations. ~\ref{Gsig} and ~\ref{GRPsig}) and effective temperatures are randomly sampled and added to the true values to produce observed values for these quantities. 

\begin{figure*}
\centering
\includegraphics[scale=0.4,trim={0 0 0 1.cm},clip]{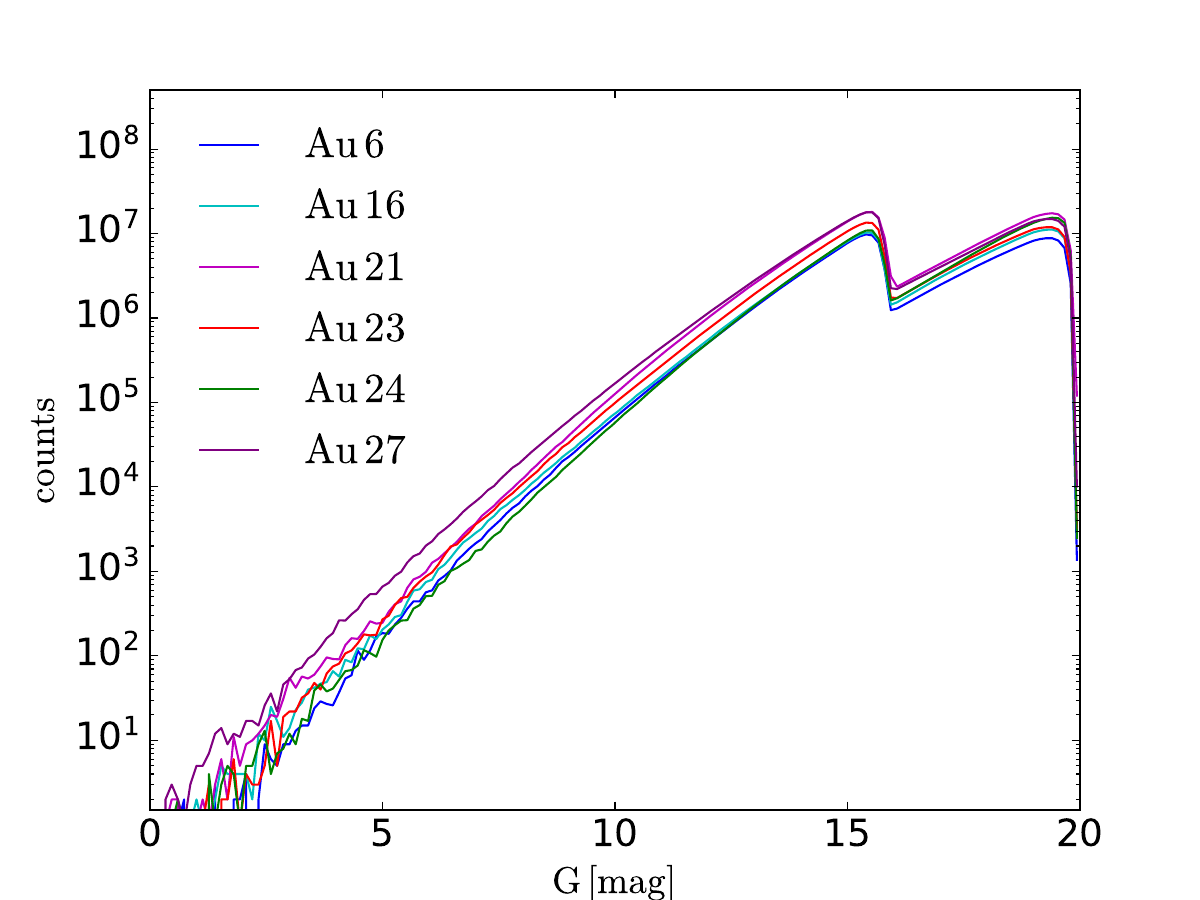}
\includegraphics[scale=0.4,trim={0 0 0 1.cm},clip]{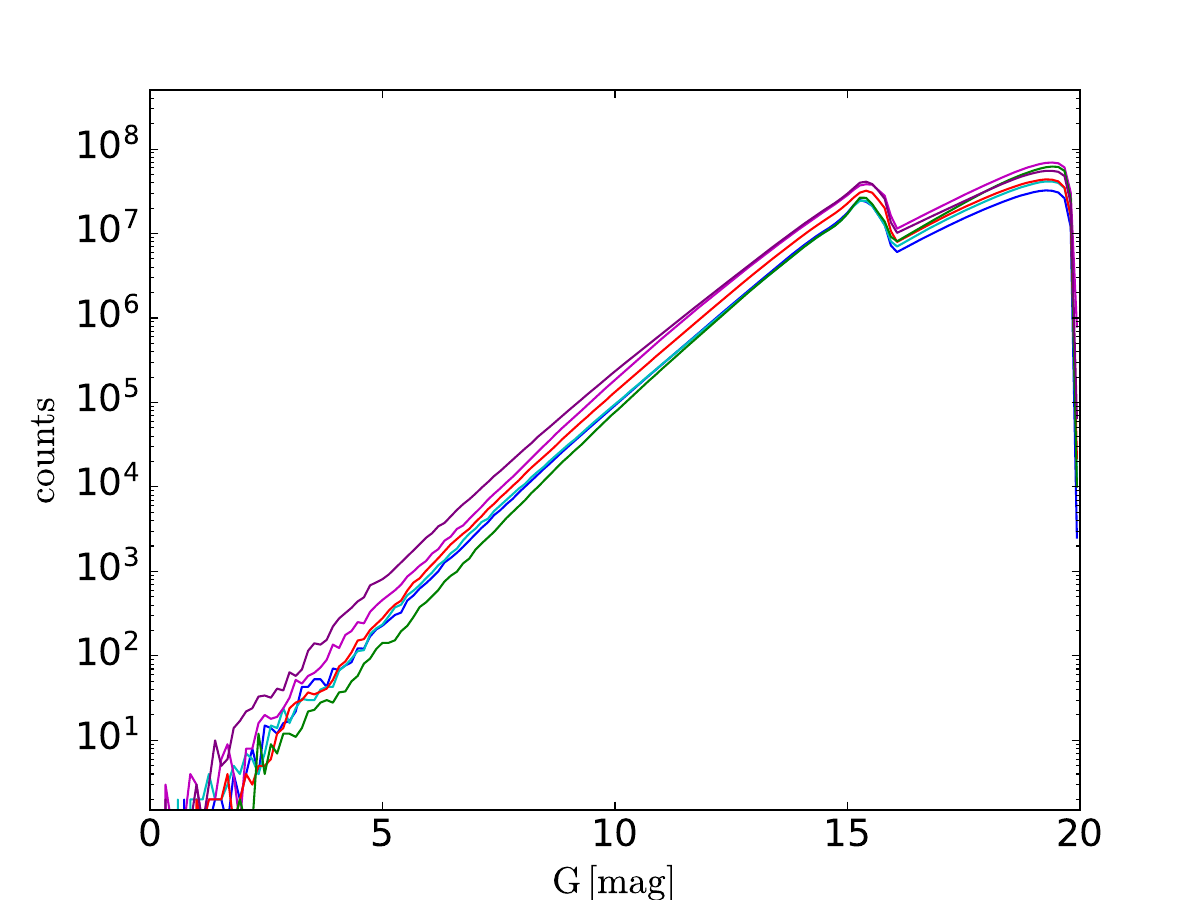}
\caption{The distribution of stars as a function of $G$-magnitude in the \Hmocks{} (left panel) and the \Dmocksnoex{} (right panel) for the default solar position of each simulation. The step at $V\sim 16$ reflects our choice to select stars with $16 < V < 20$ at latitudes $|b|>20$ degrees, whereas the stars brighter than $V=16$ are sampled with full sky coverage. The bin size is $0.1$ magnitudes.}
\label{gdist}
\end{figure*}

\begin{figure*}
    \includegraphics[scale=0.75,trim={0 0 0.7cm 0},clip]{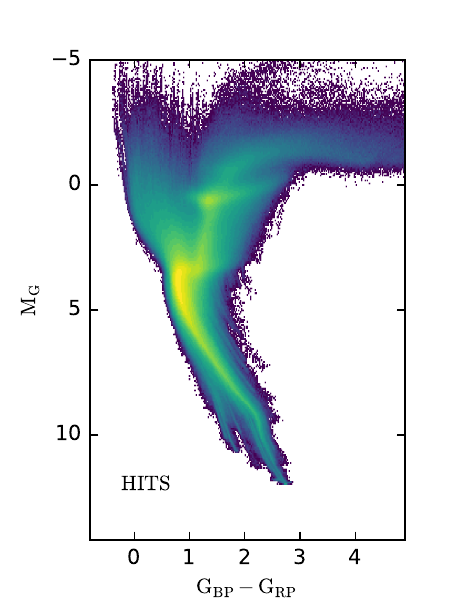}
    \includegraphics[scale=0.75,trim={1.5cm 0 0.7cm 0},clip]{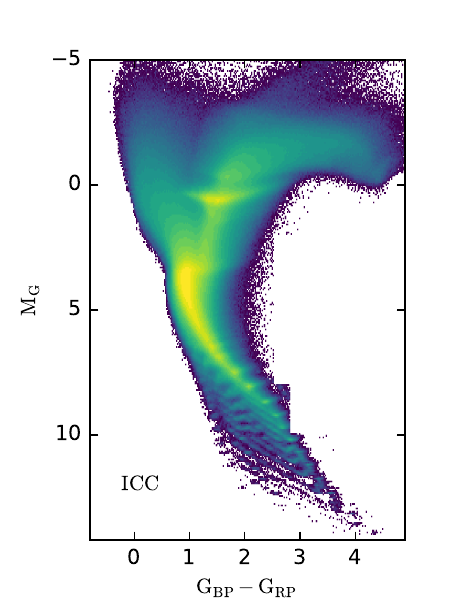}
    \includegraphics[scale=0.75,trim={1.5cm 0 0.7cm 0},clip]{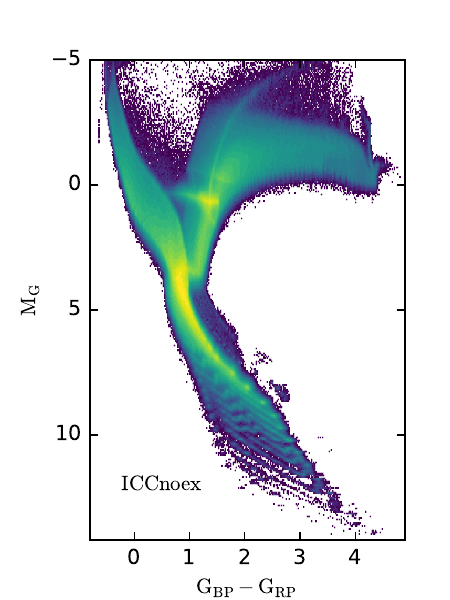}
    \includegraphics[scale=0.75,trim={1.5cm 0 0.7cm 0},clip]{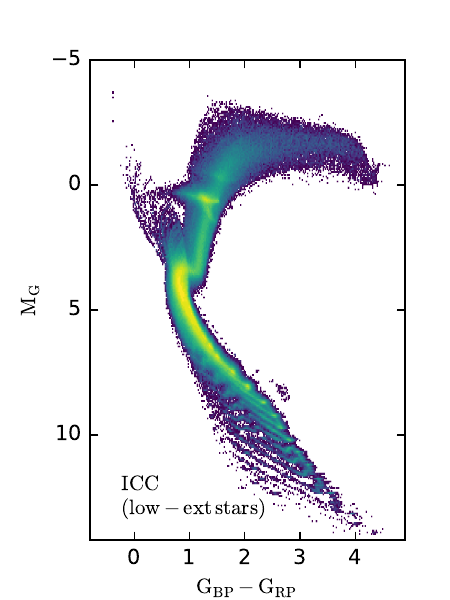}
	\caption{\emph{Gaia} colour-magnitude diagrams for mock catalogues generated for Au 24 at the default solar position. These are constructed by sampling the stellar particles taking into account the mass, age and metallicity of each particle according to the corresponding IMF. The first three panels show a \textsc{hits-mock}, \textsc{icc-mock} and \textsc{icc-mock-noex} for a subset of stars with accurate astrometry and photometry (see text for detailed selection criteria). The fourth panel shows the same as the second panel, but with an extra selection cut for stars with low-extinction.}
	\label{fig2}
\end{figure*}

\subsection{Access to Mock Catalogues}

The \Hmocks{} and \Dmocks{} presented in this paper will be made available to the community upon submission of this article. They will be available to download from the \textlcsc{AURIGA} website\footnote{\href{http://auriga.h-its.org}{\url{http://auriga.h-its.org}}} as well as the Virgo Millennium database in Durham\footnote{See \href{http://icc.dur.ac.uk/data}{\url{http://icc.dur.ac.uk/data}}}, which also allows subsets of data to be retrieved using SQL queries. In addition, snapshot particle data and gravitational potential grids will be made available at these locations. A description of the data fields and their units is given in Table.~\ref{table2}.

\begin{figure*}
	\centering
	\includegraphics[scale=0.5,trim={0 0 0 0},clip]
{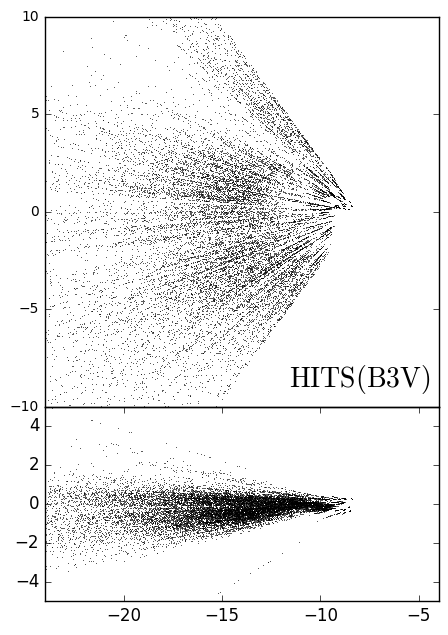}
	\includegraphics[scale=0.5,trim={0 0 0 0},clip]
{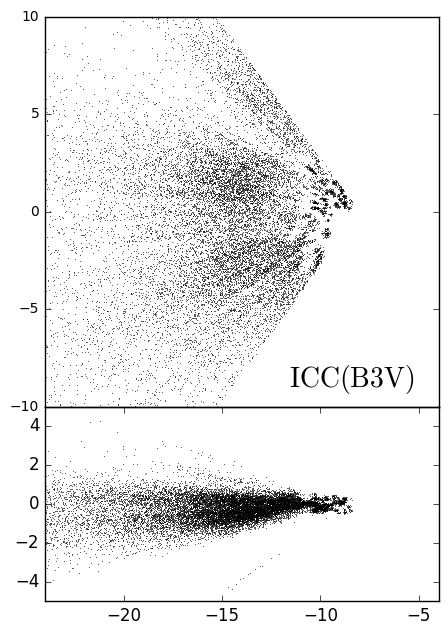}
    \includegraphics[scale=0.5,trim={0 0 0 0},clip]
{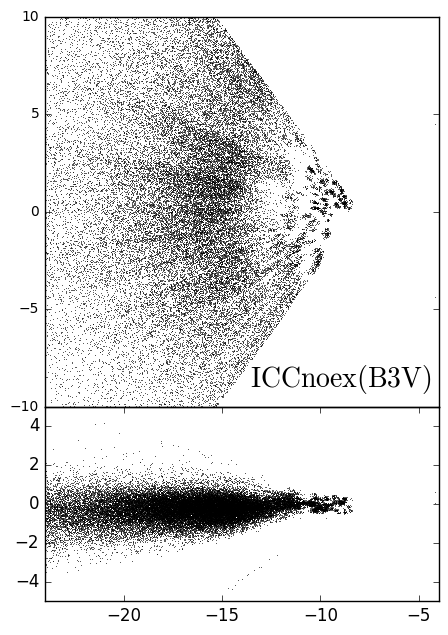}\\
	\includegraphics[scale=0.5,trim={0 0 0 0},clip]{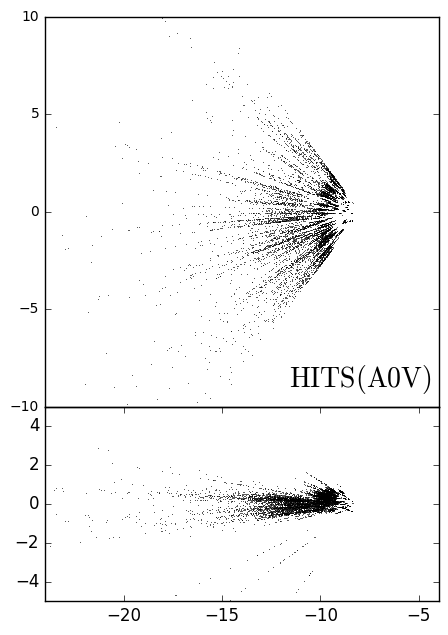}
    	\includegraphics[scale=0.5,trim={0 0 0 0},clip]{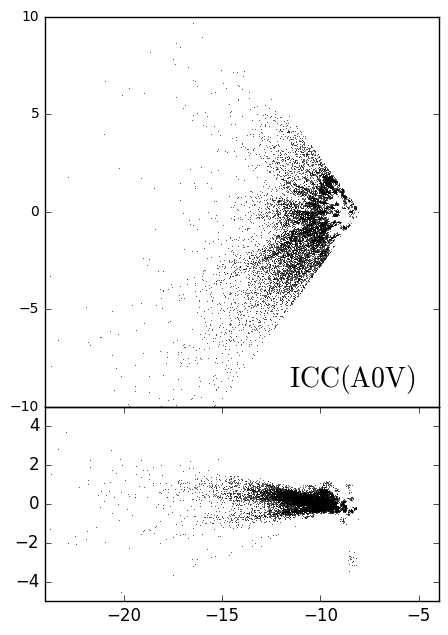}
	\includegraphics[scale=0.5,trim={0 0 0 0},clip]{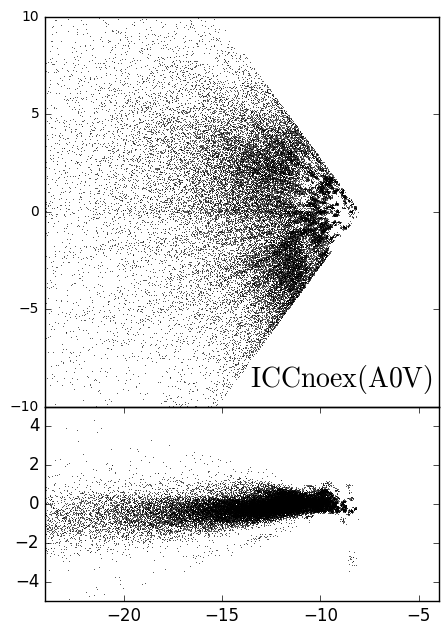}
	\caption{The face-on and edge-on distribution of $\rm B3V$ stars (top panels) and $\rm A0V$ stars (bottom panels) selected to be within a longitude of $126 < l < 234$. This is shown for in the fiducial \textlcsc{HITS-MOCK} (left panels), \textlcsc{ICC-MOCK} (middle panels) and \textlcsc{ICC-MOCKnoex} (right panels) of Au 24. The Galactic centre is located at $(X,Y,Z)=(-8, 0,0.02)$ $\rm kpc$. Note that the brighter $\rm B3V$ stars are spread over a larger portion of the disc than the $\rm A0V$ stars. The effects of dust extinction, particularly in the plane, are evident on comparison of the left and middle panels with the right panels, and are more obvious for the $\rm A0V$ stars.}
	\label{fig3}
\end{figure*}

\begin{figure*}
\centering
	\includegraphics[scale=1.1,trim={0.5cm 1.cm 0 0.7cm},clip]{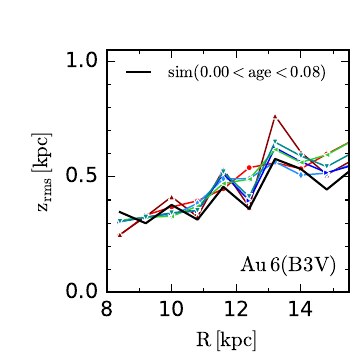}
	\includegraphics[scale=1.1,trim={1.cm 1.cm 0 0.7cm},clip]{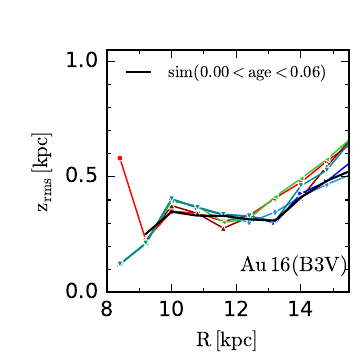}
	\includegraphics[scale=1.1,trim={1.cm 1.cm 0 0.7cm},clip]{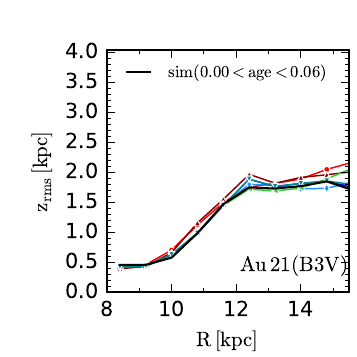}\\
	\includegraphics[scale=1.1,trim={0.5cm 0.95cm 0 0.5cm},clip]{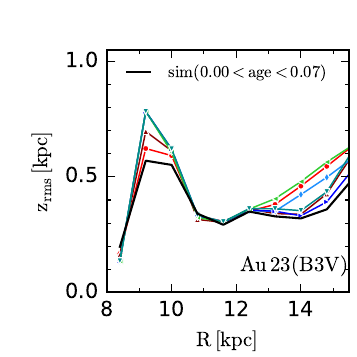}
	\includegraphics[scale=1.1,trim={1.cm 0.95cm 0 0.5cm},clip]{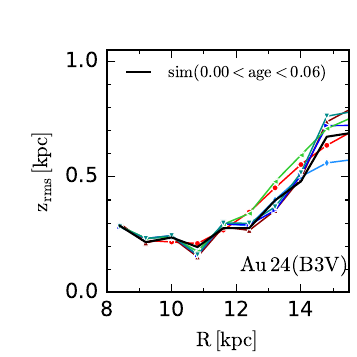}
    	\includegraphics[scale=1.1,trim={1.cm 0.95cm 0 0.5cm},clip]{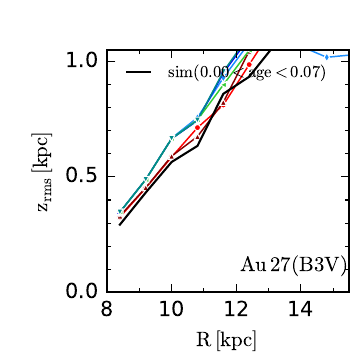}\\

    \includegraphics[scale=0.8,trim={5.5cm 1.2cm 0 1.2cm},clip]
{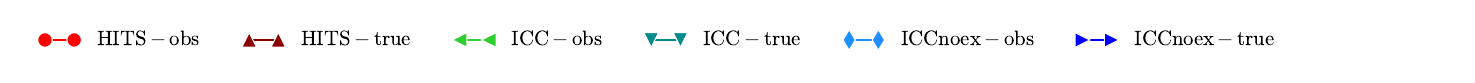}\\

    \includegraphics[scale=1.1,trim={0.5cm 1.cm 0 0.7cm},clip]{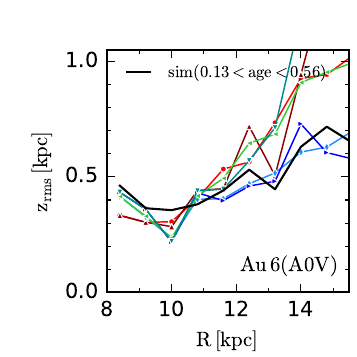}
	\includegraphics[scale=1.1,trim={1.cm 1.cm 0 0.7cm},clip]{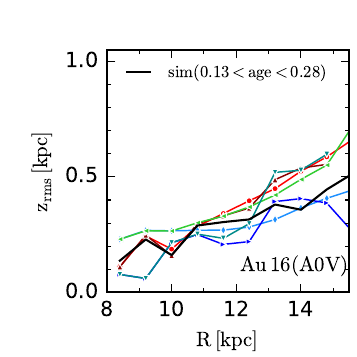}
	\includegraphics[scale=1.1,trim={1.cm 1.cm 0 0.7cm},clip]{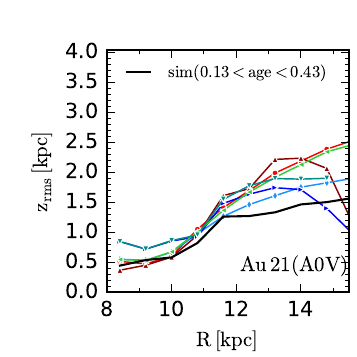}\\
	\includegraphics[scale=1.1,trim={0.5cm 0 0 0.5cm},clip]{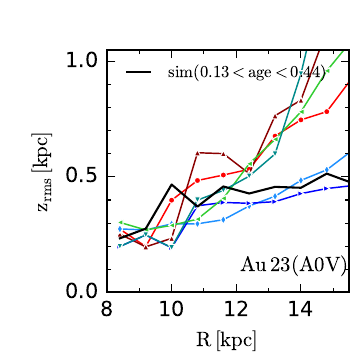}
	\includegraphics[scale=1.1,trim={1.cm 0 0 0.5cm},clip]{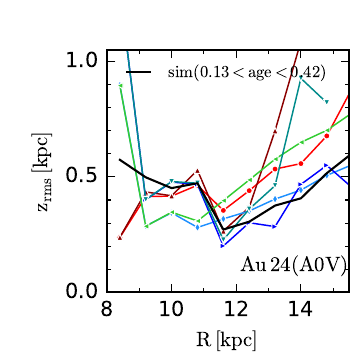}
    	\includegraphics[scale=1.1,trim={1.cm 0 0 0.5cm},clip]{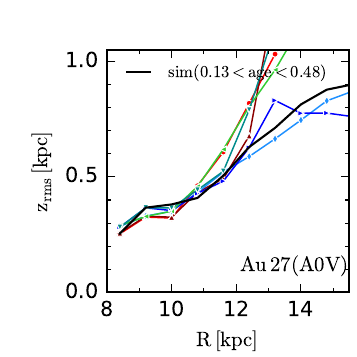}\\
    
	\caption{The root mean square vertical height as a function of radius for $\rm B3V$ dwarf stars (upper six panels) and $\rm A0V$ dwarf stars (lower six panels) from mock catalogues generated for each of the six simulations. The stars are selected in the outer disc ($126\degr < l < 234\degr$) and around narrow $M_V$ and $V-I_c$ ranges according to the values listed in \citet{PM13}. An additional cut on relative parallax error $0 < \sigma _{\pi}/\pi < 0.5$ is made. This typically results in several tens of thousands of stars that cover a large portion of the Galactic disc (see Fig.~\ref{fig3}). In each case, we show the root mean square height of: the raw simulation data for star particles of the corresponding age (black curves); the true positions of the \Hmocks{} (red) and \Dmocks{} with (green) and without (blue) extinction; the observed positions after error displacement (lighter colours) for each mock.
\newline{}
    }
	\label{fig4}
\end{figure*}

\begin{figure}
	\centering
	\includegraphics[scale=1.3,trim={0.75cm 0.1cm 0 0.2cm},clip]{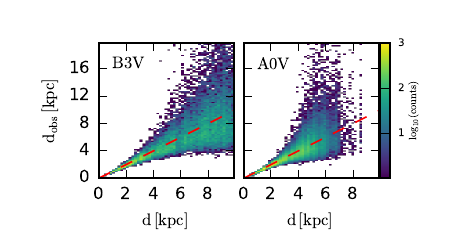}
    \vskip -.2cm
	\caption{The observed heliocentric distance (after displacement from the parent star particle) as a function of real heliocentric distance (before displacement, i.e., the parent star particle distance) for $\rm B3V$ and $\rm A0V$ stars in the default \textsc{hits-mock} of Au 24. The one-to-one relation is shown by the dashed red line.}
	\label{figerr}
\end{figure}

\begin{figure*}
	\includegraphics[scale=1.1,trim={0.5cm 1.cm 0 0.7cm},clip]{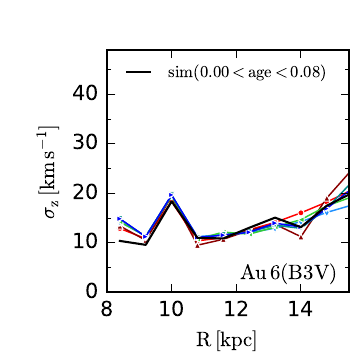}
	\includegraphics[scale=1.1,trim={1.1cm 1.cm 0 0.7cm},clip]{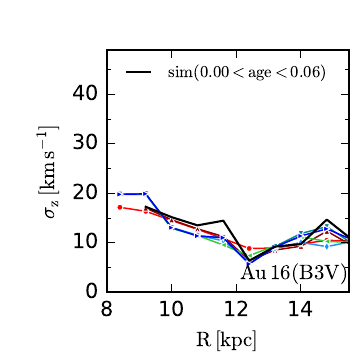}
	\includegraphics[scale=1.1,trim={1.1cm 1.cm 0 0.7cm},clip]{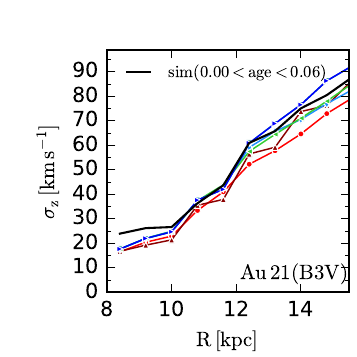}\\
	\includegraphics[scale=1.1,trim={0.5cm 0.95cm 0 0.5cm},clip]{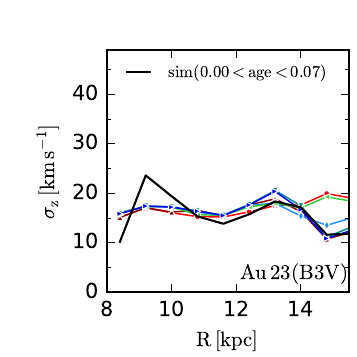}
	\includegraphics[scale=1.1,trim={1.1cm 0.95cm 0 0.5cm},clip]{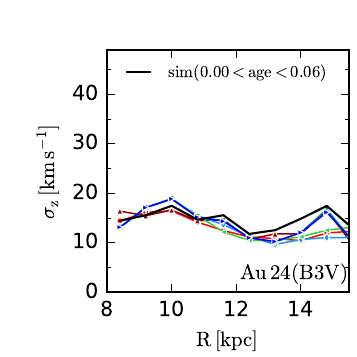}
    	\includegraphics[scale=1.1,trim={1.1cm 0.95cm 0 0.5cm},clip]{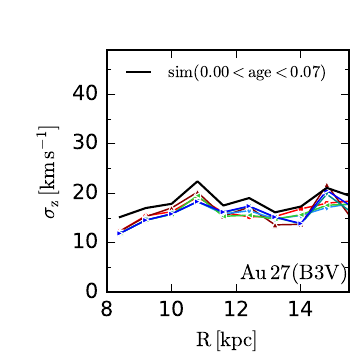}\\

    \includegraphics[scale=0.8,trim={5.5cm 1.2cm 0 1.2cm},clip]
{figures/hz_legend_wide_top}\\

    \includegraphics[scale=1.1,trim={0.5cm 1.cm 0 0.7cm},clip]{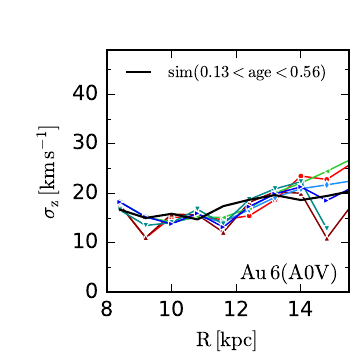}
	\includegraphics[scale=1.1,trim={1.1cm 1.cm 0 0.7cm},clip]{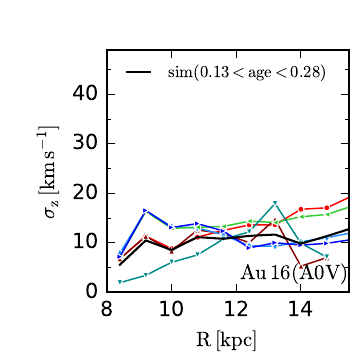}
	\includegraphics[scale=1.1,trim={1.1cm 1.cm 0 0.7cm},clip]{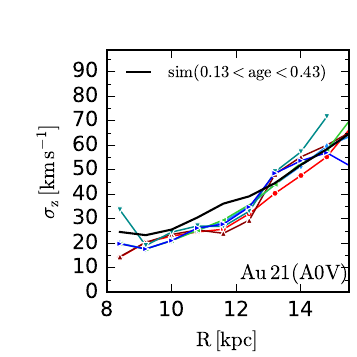}\\
	\includegraphics[scale=1.1,trim={0.5cm 0 0 0.5cm},clip]{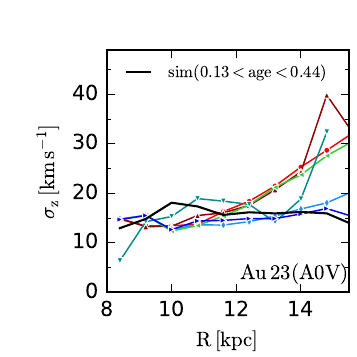}
	\includegraphics[scale=1.1,trim={1.1cm 0 0 0.5cm},clip]{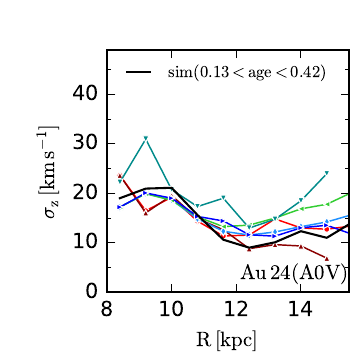}
    	\includegraphics[scale=1.1,trim={1.1cm 0 0 0.5cm},clip]{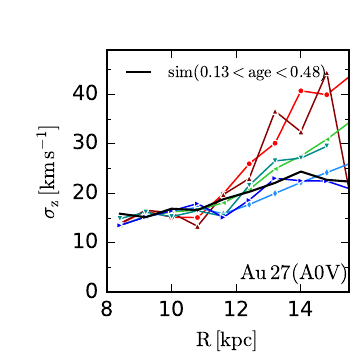}\\

	\caption{As Fig.~\ref{fig4}, but for the vertical velocity dispersion.}
	\label{fig5}
\end{figure*}

\section{The Mock Catalogues and Example Applications}
\label{results}

Fig.~\ref{fig1} shows all-sky maps of the observed mock stellar distributions in heliocentric equatorial coordinates (right ascension and declination) for one of the \Hmocks. These maps are constructed by mapping the $K$-, $G$- and $U$-band apparent magnitudes of stars to the red, green and blue colour channels of the composite image. The upper all-sky map includes all stars out to 20 kpc heliocentric distance, and clearly shows the presence of a central yellow bulge and blue disc. The blue light from nearby stars extends above the Galactic mid-plane and fades with increasing latitude. Immediately obvious is the dust obscuration in the disc mid-plane, which coincides with the disc plane and is pronounced in directions toward the bulge. In the lower all-sky map, we show all stars within 5-20 kpc distance. In this volume, the bulge and outer disc are emphasised because stars that make up these components contribute more to the map than the local disc. In turn, dust obscuration is more obvious. For clarity, Fig.~\ref{fig1} also shows the surface mass density of the mock stellar distribution in cartesian coordinates (face-on: top-right panel; edge-on: bottom-right panel). We note that the observed distribution of stars is more extended than the true distribution because the \emph{Gaia} DR2 errors can become large at large distances, which for the \Hmocks{} translates to large displacements of stars in phase space and thus to an inevitable increase in the observed phase-space domain. 

Fig.~\ref{gdist} shows the apparent $G$-magnitude distribution of stars in each of the \Hmocks{} (top panel) and \Dmocksnoex{} (bottom panel) generated from the default solar position (30 degrees behind the bar major axis). We reiterate that catalogues cover the full sky for stars with magnitudes $V<16$, whereas fainter stars with $16<V<20$ are only provided at latitudes $|b|> 20$~degrees. The lower number of stars fainter than $V=16$ reflects the $20\%$ sampling rate of these stars in the \Hmocks. These distributions do not vary significantly between the mock catalogues. We note that the $G$-magnitude distributions for the \Dmocks{} are very similar to those of the \Hmocks, therefore for brevity we omit the former.

In the first two panels of Fig.~\ref{fig2}, we show the \textsc{hits-mock} and \textsc{icc-mock} colour-magnitude diagram (CMD) for Au 24 at the default solar position. Following \citet{Gaia18d}, we selected stars with: parallax errors better than $10\%$; $G$ magnitude errors better than 0.22 mag; $G_{BP/RP}$ magnitudes better than 0.054 mag. These CMDs contain the full spectral range of main-sequence stars, and feature prominent evolutionary stages such as the main sequence turn-off and the red giant branch. The corresponding CMD of the \Dmocksnoex{} is shown in the third panel of Fig.~\ref{fig2}, and clearly illustrates the effects of reddening and extinction on comparison with the second panel: the main sequence and turn off are much sharper and bluer in the absence of dust. In the fourth panel of Fig.~\ref{fig2}, we impose an additional selection criterion for stars with little extinction ($A_0 < 0.03$) for the \textsc{icc-mock}. This enhances the clarity of the CMD features (compared to the second panel), and demonstrates that our CMDs are qualitatively similar to those presented in \citet{Gaia18d}. We note that we do not model the white dwarf sequence, which is the main difference between these simulated CMDs and \emph{Gaia} CMDs.

In the remainder of this section, we present applications of the mock data to the stellar disc and halo. We restrict ourselves to two applications, the flaring (young) stellar disc and the stellar halo spin.

\subsection{Flaring disc(s)}
\label{results:disc}

In the last years, both simulations and observations have increasingly focussed on the chemical and age structure of the stellar disc \citep[e.g.][]{ScB09,BRH12,RCK13,MCM14,HBH15,MBS17,SM17}. An interesting result of these analyses is that the outer disc of the  Milky Way is composed of sub-populations of age (and metallicity), each of which flare\footnote{The term \emph{flare} refers to an increase in scale height with increasing radius.}. This sort of flaring distribution is often seen in numerical simulations that include orbiting satellites and mergers \citep[e.g.][]{QHF93,MCM14b,MMF14} that act to preferentially dynamically heat the outer disc more than the inner disc. However, an alternative, internal mechanism that may give rise to disc flaring is the radial migration of stars from the inner disc to the outer disc: \citet{BRS15} has shown that the degree of flaring found in the APOGEE Red Clump data is consistent with theoretical predictions of the radial migration of stars under conservation of vertical action arguments \citep{Min12,SoS12,Rok12}. This finding suggests a secular dynamical origin for the flared distributions; however the origin remains to be conclusively determined and is still debated.

Although much attention has been paid to dynamical origins, there is
growing evidence that the flared distributions may be formed \emph{in
  situ} from flaring star-forming regions. \citet{GSG16} showed that a
significant amount of the vertical velocity dispersion is set at birth
from star-forming gas that becomes progressively thinner with time and
that, at a given look back time, the radial profile of the vertical
velocity dispersion of young stars ($< 1$ Gyr old) is flat,
corresponding to a flaring scale height. \citet{NYL17} showed from the
Apostle simulations \citep{SFF15,FNS16} that stars are born in flared
distributions. Moreover, these distributions do not change
significantly thereafter; they are not strongly affected by subsequent
dynamical processes. This idea that the star forming gas disc
intrinsically flares in supported also by the simple analytical
arguments put forward by \citet{BLNF18}, who demonstrated that the
vertical structure of polytropic, centrifugally supported gas discs
with flat rotation curves embedded in CDM haloes naturally
flare. Moreover, the recent controlled numerical study of
\citet{KGG17} suggests that flaring star-forming regions are required
in order to preserve a negative vertical metallicity gradient that
would otherwise become positive owing to the outward radial migration
of metal rich stars. Flaring star-forming regions have therefore
become a new and attractive way to help explain the flaring stellar
disc. 

A strong signature of an \emph{in situ} flaring disc is a flaring
distribution of very young stars ($\lesssim 300$ Myr), because radial
migration requires several dynamical times to become effective. We
therefore select young A and B dwarf stars from the mock stellar
catalogues according to the absolute $V$-band magnitude, $V-I_c$
colour and tentative ages given by \citet{PM13}, that is: ($V$,
$V-I_c$) $\sim$ ($-1.1$, $-0.192$) for $\rm B3V$ stars; and ($V$,
$V-I_c$) $\sim$ ($-1.11$, $0.004$) for $\rm A0V$ stars. These stars
are typically $\sim 0.1$ Gyr and $\sim 0.3$ Gyr old, respectively. We
select stars in the outer disc region ($126\degr < l < 234\degr$) in
order to minimise heavy midplane extinction. The distribution of
$\rm B3V$ and $\rm A0V$ stars is shown in Fig.~\ref{fig3} for
  each catalogue, and demonstrates that these stars cover a
significant portion of the outer disc, particularly in the absence of
extinction. Comparison of the left and middle panels with the
  right panels of Fig.~\ref{fig3} highlights the drastically reduced
  number of stars near the disc midplane caused by dust extinction,
  particularly for $\rm A0V$ stars. The ``fingers of God'' feature in
the distributions shown in the \Hmocks{} (left panels of
Fig.~\ref{fig3}) are caused by fluctuations in dust attenuation along
different lines of sight, and by the displacement of the true
  stellar positions along the line-of-sight due to parallax
  errors. These features are less evident in the \Dmocks{} (middle
  panels of Fig.~\ref{fig3}), because of the phase-space smoothing of
  the stars.

To make a simple estimate of the vertical thickness, we calculate
  the root mean square height (or scale height hereafter) as a
function of observed Galactocentric radius for our samples of $\rm
B3V$ and $\rm A0V$ stars, selected from mock catalogues generated for
each simulation. The radial profiles of these scale heights
  are shown in Fig.~\ref{fig4} for the default solar position of 30
  degrees behind the major axis of the bar. In addition, we compare
the scale height profiles of the B and A stars selected from each mock
catalogue with those of raw simulation star particles of equivalent
age. We show the profile given by the ``true'' positions of the
synthetic stars (before stars are displaced in phase space by errors),
and the profile given by the ``observed'' positions (after the stars
have been displaced), for all mocks. Because both the true and
observed positions in the \Dmocks{} and \Hmocks{} include
extinction, the comparison of the true and observed profiles with the
raw simulation data indicates the effects of the dust-corrected magnitude cut and
  errors separately in addition to their overall effect. 
  However, the comparison of the \Dmocks{} to the \Dmocksnoex{}
  provides a direct indication of the effects of dust extinction. All mocks
  are affected by the magnitude cut. 

The raw simulation data and the mock data exhibit flared vertical
  scale height profiles, and are, for the true mock data, in excellent
  agreement across the radial range 8 - 16 kpc for the $\rm B3V$ stars
  in all simulations. In most cases, the observed profiles are in good
  agreement with the raw simulation data, however appreciable
  deviations begin to appear at heliocentric distances greater than
  $\sim 5$ kpc for Au 16 and Au 23. This indicates that errors are
more important than extinction for $\rm B$-type dwarfs at these
distances, which is confirmed by the distance error distributions
shown in Fig.~\ref{figerr}. The agreement is worse for the $\rm A0V$
stars compared to $\rm B3V$ stars at heliocentric distances larger
than $\sim 4$ kpc. Extinction (visible in the bottom-left and -middle panels of
Fig~\ref{fig3}) seems to be mainly responsible for the deviations away
from the raw simulation data in these cases. This is reinforced by the
\Dmocksnoex{} profiles, which do not model extinction and generally
reproduce well the raw simulation data even at galactocentric radii
$\gtrsim 13$ kpc for both types of stars. We note that the scale
  height profiles for the \Hmocks{} and \Dmocks{} are very similar for
  both stellar types.

In Fig.~\ref{fig5}, we examine the radial profiles of the vertical velocity dispersion for the same stars as in Fig.~\ref{fig4}. As expected from their flaring spatial distributions, the vertical velocity dispersion is nearly constant with radius in all cases, and is, in general, well-reproduced by all mocks. Again, this is particularly true for $\rm B3V$ stars, which show minimal deviations, similar to those of their corresponding vertical scale height profiles. For $\rm A0V$ stars, the profiles are well-reproduced up to heliocentric distances of $\sim 5$ kpc, beyond which they begin to deviate noticeably in some cases. Apart from the increasing uncertainties in parallax and proper motion at these distances, an additional inaccuracy that contributes to the observed deviations is the lack of a radial velocity component in \emph{Gaia} DR2 for these stars, although it is likely a minimal contribution for this application because radial velocities are almost perpendicular to the vertical velocity field at these low latitudes. The \Dmocksnoex{} are able to reproduce the vertical velocity dispersion for both $\rm B3V$ and $\rm A0V$ stars very well, and tend to bear out a more accurate representation of the dispersion at larger radii where extinction begins to affect the \Hmocks{} measurement of the $\rm A0V$ stars. Again, we note that the vertical velocity dispersion profiles for the \Hmocks{} and \Dmocks{} are very similar for both stellar types.

The results presented in Figs.~\ref{fig4} and ~\ref{fig5} demonstrate that, for \emph{Gaia} DR2, $\rm BV$ and $\rm AV$ stars are reliable tracers for the very young stellar disc and, by extension, the distribution of star-forming regions; the intrinsic flaring of the star-forming gas disc is captured by these dwarf stars in both position and velocity space. It is worth to note that for subsequent data releases the reliability of these tracers will improve: the ability to trace the young disc will extend to the outer reaches of the disc and the warp beyond. 

\begin{figure}
	\centering
	\includegraphics[scale=0.6,trim={0 0 0 0},clip]{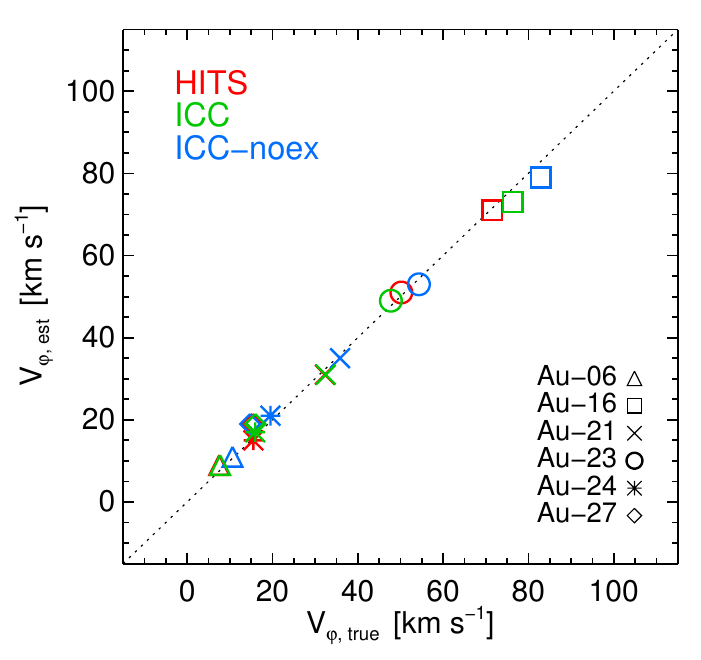}
	\caption{Estimated mean $V_{\phi}$, based on the method of \citet{deason17} from 5D data of a random sample of $70,000$ HB halo stars in \Hmocks{} (red), \Dmocks{} (blue), and \Dmocksnoex{} (green), versus the true mean $V_{\phi}$ calculated from the 6D phase-space information of the same samples. The different symbol types represent the six \textlcsc{AURIGA} simulations for which the mock catalogues are created as indicated in the legend.}
	\label{spin1}
\end{figure}

\begin{figure*}
	\centering
	\includegraphics[scale=0.8,trim={0 0 0 0},clip]{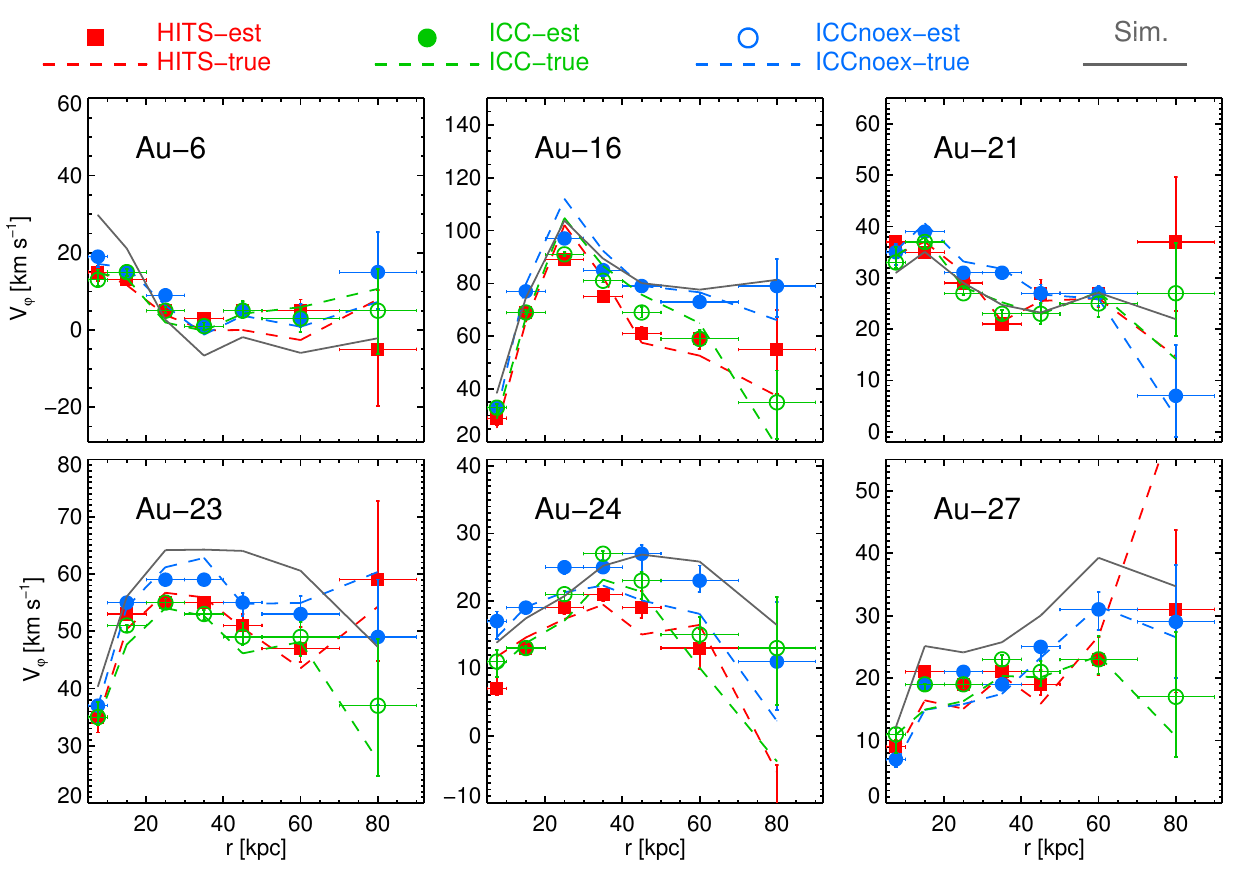}
	\caption{Estimated $V_{\phi}$ from 5D data of HB halo stars at
          different galactocentric radial bins for
            \textlcsc{AURIGA} galaxies compared to the true $V_{\phi}$
            of the same samples. Red, green and blue colours
            correspond to \Hmocks{}, \Dmocks{}, and \Dmocksnoex{},
            respectively. Different symbol types represent estimated
            values, while the true values are shown with dashed
            lines. The horizontal error bars illustrate the size of
            the radial bins, while the vertical error bars show the
            error in the mean values. The profile of raw simulated
            star particles, with a spatial cut similar to that in the
            mock catalogues and age $>10$~Gyr, are shown with a solid
            grey line, for reference. } 
	\label{spin2}
\end{figure*}

\subsection{Stellar halo rotation}
\label{results:halo}

The spin of the Milky Way stellar halo is directly related to its merger history. To first order, the stellar halo rotation represents the net angular momentum of all of the Galaxy's past accretion events. Moreover, the presence of \textit{in situ} halo stars, which are formed in the Galactic disc and later ``kicked out" into the halo due to merger events, can lead to disc-like kinematics in the stellar halo (i.e. net prograde rotation in the same sense as the disc, see e.g. \citealt{mccarthy12,cooper15a,pillepich15}). Thus, by measuring the net spin of the stellar halo we are probing the global accretion history of the Galaxy. In addition, we can gain further insight by measuring the halo rotation as a function of metallicity, Galactocentric radius, and position on the sky (see e.g. \citealt{carollo07, carollo10, deason11, kafle13, hattori13}).

Previous works attempting to measure the net spin of the halo have aimed to tease out the rotation signal using line-of-sight velocities from large spectroscopic samples of halo tracers (e.g. \citealt{sirko04, deason11}), this limitation to one velocity component is particularly troublesome for measuring rotation; at large distances  the line-of-sight velocity is essentially the radial velocity component, and there is little, or no, constraint on the tangential velocity of halo stars.  Prior to the \textit{Gaia} era, reliable proper motion measures of distant halo stars were scarce, with ground-based samples subject to large systematic uncertainties (e.g. \citealt{gould04}), and space-based samples limited to very small areas of the sky (\citealt{deason13, cunningham16}). 

Now, in the era of \textit{Gaia} DR2, we have access to \textit{all sky} proper motion measurements, with well-defined systematic and statistical error distributions. A prelude to the astrometric breakthrough from DR2 was presented in \cite{deason17}, who used a proper motion catalogue constructed from SDSS images and \textit{Gaia} DR1 to measure the net spin of the halo. The main drawback of the SDSS-\textit{Gaia} proper motion catalogue is the constraint to the SDSS sky coverage, and the limited number of known halo tracers that could be used.

In this Section, we use the mock catalogues to illustrate how \textit{Gaia} DR2 astrometry can be used to measure the net spin of the stellar halo out to 100 kpc. The \textit{Gaia} spacecraft is expected to observe $N \sim 70,000$ Galactic halo RR Lyrae stars out to $\sim 100$ kpc (\citealt{clem16}). These old, metal-poor stars are approximate standard candles, and their distances can typically be measured with accuracies of less than 5 percent (see e.g. \citealt{iorio18}). Here, we randomly sample $N \sim 70,000$ ``old" (age $> 9$ Gyr) horizontal branch (HB) stars in the \textlcsc{AURIGA} haloes with $0<B-V<0.7$ and $0.2<M_V<1.2$. This selection was chosen to approximately mimic the all-sky RRL catalogues that will be released with \textit{Gaia} DR2. To select halo stars, we include stars between 5 and 100~kpc from the Galactic centre, and $|b| > 20$ deg above/below the disc plane, and height $|z|>4$ kpc. We do not include distance uncertainties in the analysis (but note that including $\sim 5\%$ distance errors makes little difference to our results), and assume that while proper motions measurements are available from \textit{Gaia} DR2, there are no line-of-sight velocities.

In order to measure the halo rotation, we employ the same method introduced by \cite{deason17} to measure rotation with 5D data. In brief, we adopt a 3D velocity ellipsoid aligned in spherical coordinates, which assumes Gaussian velocity distributions and allows for net streaming in the $v_\phi$ component. A likelihood analysis is used to determine the best-fit $\langle v_\phi \rangle$ value. For more details on this method we refer the reader to \cite{deason17}. 

Fig.~\ref{spin1} shows the resulting mean rotation of stars in the radial range $r= 5-50$~kpc for 6 \textlcsc{AURIGA} haloes in \Hmocks{}, \Dmocks{},  \Dmocksnoex{}. The estimated $\langle v_\phi \rangle$ using the method of Deason et al., $ v_{\phi, \rm est}$, is in very good agreement with the true value for the same samples of stars ($ v_{\phi, \rm true}$). The errors on the mean values are smaller than the size of the symbols and therefore are omitted. The $v_{\phi, \rm true}$ values differ between the two mocks because different isochrones and IMFs are used, and thus our criteria for selecting old HB stars yield different subsets of stars. 
This point is important and illustrates that different subsets of old stars can have different rotation signals. We plan to investigate this further in a follow-up paper.

In Fig.~\ref{spin2} we show the estimated and true $v_{\phi}$ of our
sample of old halo stars at different radii for all the
  mocks. The method of \cite{deason17} works very well at all radii
to recover the actual spin of our samples of stars. It is remarkable
that even at distances as large as 100 kpc, where \gaia\ proper motion
errors are large and the number of stars is relatively small, one can
recover the spin of the halo stars within $2-\sigma$. The spin
  profiles of raw star particles in the simulation are shown in this
  figure, as a reference, with a grey solid line. The particles are
  chosen to have the same spatial cut as the mock stars, but with age
  older than $10$~Gyr; this is roughly equivalent to the
  color-magnitude criteria we adopted to select HB stars. Slight
  differences between the profiles from mocks and simulations are
  expected as the sample of stars are different.      

We note that Au-6 is the closest example to the MW according to 
  halo spin, which was shown by \cite{deason17} to be in the range
  $\sim 0-20~\mathrm{km\,s^{-1}}$ at galactocentric radii smaller than
  $50$~kpc.

\section{Discussion and Conclusions}
\label{sec4}

We have presented several mock Milky Way stellar catalogues designed
to match the selection criteria, volume and observable properties
(including uncertainties) of stars with $V<16$ mag and $V<20$ mag at
$|b|>20$ degrees that will be provided by the \emph{Gaia} data release
2. We employed two methods to calculate two sets of mock catalogues at
four solar-like positions (equidistant in Galactic azimuth) from
several high-resolution cosmological-zoom simulations: the \Hmocks{}
\citep[generated with a parallelised version of
\textlcsc{SNAPDRAGONS},][]{HKG15}; and the \Dmocks{} using the
\citet{LWC15} method, which distributes stars in phase space by
conserving the phase-space volume associated with each simulation
stellar particle. Both sets of mocks take into account a simple
  dust extinction model; however we produced also a full set of
  catalogues without extinction: \Dmocksnoex. All mock catalogues
provide \emph{Gaia} DR2 data products: six-dimensional phase space
information, magnitudes in the \emph{Gaia} $G$-, $G_{RP}$- and
$G_{BP}$-photometric bands, effective temperature and dust extinction
values, and include uncertainty estimates for the \emph{Gaia} DR2
astrometric, photometric and spectroscopic quantities. In addition,
the catalogues provide the age, metallicity, mass, stellar surface
gravity, gravitational potential and photometry for non-\emph{Gaia}
bands for each of the generated stars. The catalogues are available
online at both the \textlcsc{AURIGA} website and at the Durham
database centre, the latter of which provides a query-based system to
retrieve subsets of data. Gravitational potential grids and raw
snapshot data for a subset of simulations are available for download
at the \textlcsc{AURIGA} website.

\subsection{Limitations}
\label{sec:limitations}

While the mock catalogues presented in this paper have great potential
for helping to understand the formation of structure in our Milky Way
in tandem with \emph{Gaia} data, there are, of course, some
limitations to each of the methods used to generate the catalogues.

\paragraph*{Limitations of both methods:}
Neither method guarantees that the positions and velocities of mock stars are
consistent with bound orbits in the simulation potential. Caution and
careful sample selection based on filtering out stars with large
errors should be followed for any applications that require precise
correspondence between the motions of stars and their local
gravitational potential, or that are sensitive to a small number of
stars with very high velocities. 

An important limitation worth bearing in mind is that the
  simulations have finite resolution. Even though the Auriga project
  includes some of the highest resolution simulations of Milky Way
  analogues performed so far, a star particle represents a single
  stellar population of a few thousand solar masses. ``Exploding''
  these stellar particles into individual stars does not increase the
  resolution but allows a denser sampling of the phase space occupied
  by the original particles.

\paragraph*{\Dmocks{} limitations:}	
\citet{LWC15} describe how the parameters entering the phase-space sampling step in the construction of the \Dmocks{} were tuned to the values given in section~\ref{sec:phase_space_sampling}. This tuning sought to balance a sufficiently significant degree of expansion of stars away from their parent simulation particles against the preservation of coherent phase space structures, such as tidal streams, and the suppression of bias in the bulk kinematics of the stellar halo. \citet{LWC15} studied collisionless $N$-body simulations, so the same approach and parameters are not guaranteed to be optimal for the massive, coherent baryonic discs in hydrodynamical simulations like Auriga. In particular, when we compute scale lengths for a star particle formed in situ in the main galaxy, we treat \textit{all} the other in situ stars as its potential phase space neighbours. This may be a substantial approximation, because the set of all in situ particles comprises many different stellar populations that originate in different regions of phase space at different times. Treating all these  as potential neighbours of one another can lead to `cross-talk' between distinct dynamical structures, a form of over-smoothing (which is mitigated in the case of accreted halo stars by only considering particles from the same progenitor satellite as potential neighbours). For example, the scale height and vertical velocity dispersion of young, kinematically cold stars in the disc may (in principle) be inflated if neighbours from a kinemtically hotter bulk population dominate the kernels associated with their parent particles. However, in practice, we see no evidence of any significant bias in the analyses of young disk stars we present here. The possibility of artifacts arising from the phase space sampling procedure should be kept in mind nevertheless, especially in applications that probe phase space structure on very small scales.

\paragraph*{\Hmocks{} limitations:}
The \Hmocks{} do not include a phase space sampling step, i.e. the generated stars are not interpolated in phase space, before adding \emph{Gaia} DR2 errors to the particle phase space coordinates. This may create artefacts for structures that are ``long'' and ``thin'', such as the great circle stream, that arise from the displacement of stars along the line-of-sight with very similar celestial coordinates. Furthermore, the observed positions generated by displacing the coordinates of the parent star particle can be spread over large ranges for particles beyond $\sim 10$ kpc heliocentric distances, where the errors become large. This means that using parallax distances for some halo stars directly can become unreliable, and more sophisticated approaches, such as the one used in this paper, are required.\newline

We conclude that the \Dmocks{} are perhaps better suited than the \Hmocks{} for studying streams, other inhomogeneities and debris in the stellar halo, owing to the refined phase space sampling. Both sets of mocks include a model for dust extinction that allows the user to make quick assessments of how dust affects {\it Gaia} observables, which is particularly important for the stellar disc. Conversely, the \Dmocksnoex{} provide the user freedom to add any dust model to the data. The mock catalogues presented in this paper are therefore complementary and provide a wide scope for assessing the biases and capabilities of the {\it Gaia} DR2.

We note that the codes used to generate these mock catalogues may be improved in the future, in which case the mock catalogues on our public database will be updated accordingly. We urge users to refer back to the database whenever a new application is considered.

\subsection{Applications}

As a first science application of the mocks, we analysed the vertical structure of the young stellar disc and found that all simulations showed a flaring vertical scale height profile with a consistently flat vertical velocity dispersion profile. We verified that $\rm B3V$ and $\rm A0V$ stars in the outer disc selected from the mock catalogues reproduce these trends; young B and A dwarf star data in DR2 should be reliable tracers of the young stellar disc. If in the \emph{Gaia} DR2 data these tracers exhibit flaring profiles, this will constitute evidence for flaring star-forming regions, and perhaps indicate that radial migration and dynamical heating from satellite perturbations are not the principal drivers of the flaring mono-abundance populations found in other Galactic surveys \citep{BRS15,MBS17}.

We also applied the method of \citet{deason17} to samples of old horizontal branch halo stars in the mock catalogues to estimate the mean rotation of \textlcsc{AURIGA} stellar haloes based on 5D phase-space information. We find excellent agreement between the estimated mean rotation velocity and the true values, even at galactocentric distances as large as $100$~kpc. The results show that accurate distance measurements combined with proper motions from \gaia, can reliably predict the mean rotation of halo stars. Obtaining an accurate estimate of the spin of the distant MW stellar halo is therefore extremely promising using the tens of thousands of RR Lyrae stars that \gaia\ will provide.

The mock catalogues presented in this paper are the first such catalogues generated from \emph{ab initio} high-resolution $\Lambda$CDM galaxy formation simulations; they  offer a novel perspective of the Milky Way and may be used for a variety of applications. In particular, they provide  a testbed for the design and evaluation of Galaxy modelling methods in a realistic cosmological setting, a means to gauge the limitations and biases of \emph{Gaia} DR2 and to link observations to theoretical predictions, encapsulated in the simulations, enabling robust inferences to be made about the multitude of galaxy formation processes that shaped the Milky Way.

\section*{acknowledgements}
The authors thank the referee for a constructive and helpful report that led to the improvement of the manuscript. RG would like to thank Daisuke Kawata for many useful discussions. RG and VS acknowledge support by the DFG Research Centre SFB-881 `The Milky Way System' through project A1. JASH is supported by a Dunlap Fellowship at the Dunlap Institute for Astronomy \& Astrophysics, funded through an endowment established by the Dunlap family and the University of Toronto. AD is supported by a Royal Society University Research Fellowship. AF is supported by a European Union COFUND/Durham Junior Research Fellowship (under EU grant agreement no. 609412). MC was supported by Science and Technology Facilities Council (STFC) [ST/P000541/1]. FAG acknowledges support from Fondecyt Regular 1181264. This work has also been supported by the European Research Council under ERC-StG grant EXAGAL- 308037 and the Klaus Tschira Foundation. Part of the simulations of this paper used the SuperMUC system at the Leibniz Computing Centre, Garching, under the project PR85JE of the Gauss Centre for Supercomputing. This work used the DiRAC Data Centric system at Durham University, operated by the Institute for Computational Cosmology on behalf of the STFC DiRAC HPC Facility \href{www.dirac.ac.uk}{\url{www.dirac.ac.uk}}. This equipment was funded by BIS National E-infrastructure capital grant ST/K00042X/1, STFC capital grant ST/H008519/1, and STFC DiRAC Operations grant ST/K003267/1 and Durham University. DiRAC is part of the National E-Infrastructure.

\bibliographystyle{mnras}
\bibliography{main}

\appendix

\section{Fields and units of the mock catalogues}
\label{app1}

The mock catalogues are in hdf5 file format, and can be downloaded in their entirety or queried through an \texttt{SQL} database system at \href{http://data.cosma.dur.ac.uk:8080/gaia-mocks/}{\url{http://data.cosma.dur.ac.uk:8080/gaia-mocks/}}. The data products and units are listed in each catalogue file, and are listed in Table.~\ref{table2}. A basic \textlcsc{PYTHON} script to read the mock data and perform coordinate transformations, and example \texttt{SQL} queries are provided on the \textlcsc{AURIGA} website \href{http://auriga.h-its.org}{\url{http://auriga.h-its.org}}.

\begin{table*}
\centering
\caption{Description of the data products and their units of the mock catalogues. Quantities denoted $^a$ and $^b$ are present in the \Hmocks{} and \Dmocks{} only, respectively. For clarity, $\alpha$, $\delta$ and $\pi$ are the right ascension, declination and parallax, respectively, and $\mu^*_{\alpha}$, $\mu _{\delta}$ and $v_r$ are the proper motion right ascension in true arc ($\mu^*_{\alpha}=\mu_{\alpha} cos(\delta)$), the proper motion declination and heliocentric radial velocities, respectively.}
\begin{tabular}{c c c}
\hline
Catalogue field name & Units & Description \\
\hline                                                                     
\texttt{AccretedFlag} & - & equal to either (-1, 0, 1) for (in-situ, accreted, in existing subhalo)\\ 
\texttt{Age} & gigayears & the look back time at which the parent star particle is born \\
\texttt{EffectiveTemperature} & Kelvin & the true effective temperature of the synthetic star \\    
\texttt{EffectiveTemperatureError} & Kelvin & the error in effective temperature of the synthetic star \\    
\texttt{EffectiveTemperatureObs} & Kelvin & the observed effective temperature of the synthetic star \\    
$^a$\texttt{Extinction31} & magnitudes & $V$-band extinction value \\           
\texttt{GBmagnitude} & magnitudes & true \emph{Gaia} blue $G_{\rm B}$-band luminosity \\ 
\texttt{GBmagnitudeError} & magnitudes & error in \emph{Gaia} blue $G_{\rm B}$-band luminosity \\        
\texttt{GBmagnitudeObs} & magnitudes & observed \emph{Gaia} blue $G_{\rm B}$-band luminosity \\          
\texttt{GRmagnitude} & magnitudes & true \emph{Gaia} red $G_{\rm R}$-band luminosity \\             
\texttt{GRmagnitudeError} & magnitudes & error in \emph{Gaia} red $G_{\rm R}$-band luminosity \\         
\texttt{GRmagnitudeObs} & magnitudes & observed \emph{Gaia} red $G_{\rm R}$-band luminosity \\           
\texttt{Gmagnitude} & magnitudes & true \emph{Gaia} white light $G$-band luminosity \\ 
\texttt{GmagnitudeError} & magnitudes & error in \emph{Gaia} white light $G$-band luminosity \\         
\texttt{GmagnitudeObs} & magnitudes & observed \emph{Gaia} white light $G$-band luminosity \\           
\texttt{GravPotential} & $\rm km^2\,s^{-2}$ & gravitational potential of the parent star particle \\           
\texttt{HCoordinateErrors} & (radians, radians, arcsec) & 2D array of errors in ($\alpha$, $\delta$, $\pi$) \\ 
\texttt{HCoordinates} & (radians, radians, arcsec) & 2D array of true ($\alpha$, $\delta$, $\pi$) \\ 
\texttt{HCoordinatesObs} & (radians, radians, arcsec) & 2D array of observed ($\alpha$, $\delta$, $\pi$) \\   
\texttt{HVelocities} & ($\rm arcsec \, yr^{-1}$, $\rm arcsec \, yr^{-1}$, $\rm km\, s^{-1}$) & 2D array of true ($\mu^* _{\rm \alpha}$, $\mu _{\rm \delta}$, $v_r$)\\   
\texttt{HVelocitiesObs} & ($\rm arcsec \, yr^{-1}$, $\rm arcsec \, yr^{-1}$, $\rm km\, s^{-1}$) & 2D array of observed ($\mu^* _{\rm \alpha}$, $\mu _{\rm \delta}$, $v_r$)\\   
\texttt{HVelocityErrors} & ($\rm arcsec \, yr^{-1}$, $\rm arcsec \, yr^{-1}$, $\rm km\, s^{-1}$) & 2D array of errors in ($\mu^* _{\rm \alpha}$, $\mu _{\rm \delta}$, $v_r$)\\          
\texttt{IabsMagnitude} & magnitudes & $I$-band absolute magnitude \\            
\texttt{Magnitudes} & (magnitudes)$\times 8$ & 2D array of apparent magnitudes in the (U, B, R, J, H, K, V, I) bands \\
$^b$\texttt{InitialMass} & solar masses & mass of the star when it was born (before mass loss occurs) \\
\texttt{Mass} & Solar masses & mass of the star \\                    
\texttt{Metallicity} & - & metallicity of the star \\             
\texttt{ParticleID} & - & unique ID of the parent particle \\              
\texttt{SurfaceGravity} & log & logarithm of the true surface gravity of the star\\   
\texttt{SurfaceGravityError} & log & logarithm of the error in surface gravity of the star\\     
\texttt{SurfaceGravityObs} & log & logarithm of the observed surface gravity of the star\\       
\texttt{VabsMagnitude} & magnitudes & $V$-band absolute magnitude \\    
\hline
\end{tabular}
\label{table2}
\end{table*}

\label{lastpage}

\end{document}